\documentclass[sn-aps, iicol]{sn-jnl}% American Physical Society (APS) Reference Style
\usepackage{physics}%for ket notation

\usepackage{dcolumn}% Align table columns on decimal point
\usepackage{bm}% bold math

\newcommand{\onlinecite}[1]{\hspace{-1 ex} \nocite{#1}\citenum{#1}}

\jyear{2021}%

\theoremstyle{thmstyleone}%
\theoremstyle{thmstyletwo}%

\theoremstyle{thmstylethree}%

\begin{document}

\title[Qubit-oscillator relationships in the open quantum Rabi model: the role of dissipation]{Qubit-oscillator relationships in the open quantum Rabi model: the role of dissipation}


\author*[1]{\fnm{G.} \sur{Di Bello}} \email{grazia.dibello@unina.it}

\author[2]{\fnm{L. M.} \sur{Cangemi}} %\email{lorismaria.cangemi@gmail.com}
\equalcont{These authors contributed equally to this work. }

\author[3,4]{\fnm{V.} \sur{Cataudella}} %\email{vittorio.cataudella@gmail.com}
\equalcont{These authors contributed equally to this work. }

\author[3,4]{\fnm{G.} \sur{De Filippis}} %\email{giuliode70@gmail.com}
\equalcont{These authors contributed equally to this work. }

\author[5]{\fnm{A.} \sur{Nocera}} %\email{alberto.nocera@ubc.ca}
\equalcont{These authors contributed equally to this work. }

\author*[3]{\fnm{C. A.} \sur{Perroni} } \email{carmine.perroni@unina.it}
\equalcont{These authors contributed equally to this work.} 

\affil*[1]{\orgdiv{Dip. di Fisica E. Pancini}, \orgname{Università di Napoli Federico II}, \orgaddress{\city{Napoli}, \postcode{80126}, \country{Italy}}}   

\affil[2]{\orgdiv{Department of Chemistry}, \orgname{Bar-Ilan University}, \orgaddress{\city{Ramat-Gan}, \postcode{52900}, \country{Israel}}}   

\affil[3]{\orgdiv{SPIN-CNR and Dip. di Fisica E. Pancini}, \orgname{Università di Napoli Federico II}, \orgaddress{\city{Napoli}, \postcode{80126}, \country{Italy}}}   

\affil[4]{\orgdiv{INFN, Sezione di Napoli, Complesso Universitario di Monte S. Angelo}, \orgname{Università di Napoli Federico II}, 
\orgaddress{\city{Napoli}, \postcode{80126}, \state{Italy}}}   

\affil[5]{\orgdiv{Department of Physics and Astronomy and Stewart Blusson Quantum Matter Institute}, \orgname{University of British Columbia}, \orgaddress{\city{Vancouver}, \postcode{V6T 1Z1}, \state{B.C.}, \country{Canada}}}   


\abstract{Using a dissipative quantum Rabi model, we study the dynamics of a slow qubit coupled to a fast quantum harmonic oscillator interacting with a bosonic bath from weak to strong and ultra-strong coupling regimes. Solving the quantum Heisenberg equations of motion, perturbative in the internal coupling between qubit and oscillator, we derive functional relationships directly linking the qubit coordinates in the Bloch sphere to oscillator observables. We then perform accurate time-dependent Matrix Product State simulations, and compare our results both with the analytical solutions of the Heisenberg equations of motion, and with numerical solutions of a Lindblad master equation, perturbative in the external coupling between oscillator and environment. 
Indeed, we show that, up to the strong coupling regime, the qubit state accurately fulfills the derived functional relationships. We analyse in detail the case of a qubit starting with generic coordinates on the Bloch sphere of which we evaluate the three components of the Bloch vector through the averages of oscillator observables. 
Interestingly, a weak to intermediate oscillator coupling to the bath is able to simplify the Bloch vector evaluation since qubit-oscillator relationships are more immediate. Moreover, by monitoring the qubit fidelity with respect to free limit, we find the parameter regime where the combined effect of internal and external couplings is able to hinder the reliable evaluation of the qubit Bloch vector.
Finally, in the ultra-strong coupling regime, non-Markovian effects become robust and the dynamics of qubit and oscillator are inextricably entangled making the qubit Bloch vector evaluation difficult.} 

%\keywords{keyword1, Keyword2, Keyword3, Keyword4}



\maketitle

\thispagestyle{plain} 

\begin{twocolumn}
\section{Introduction}\label{sec:intro}
The interaction between matter and electromagnetic fields is one of the most fundamental processes occurring in nature. This interaction, specified to quantized electromagnetic fields, is the central focus of investigation in cavity quantum electrodynamics (QED) and in circuit QED, with artificial atoms and on-chip resonators. This research field is really important for both atomic physics and quantum optics \cite{gu2017microwave}. The quantum Rabi model \cite{rabi1936process} describes the most used setup, in which a two-level system (qubit) is coupled to a harmonic oscillator. This model gives a mathematical framework for various quantum phenomena, inter alia, nonclassical-state generation, quantum state transfer and topological physics using photons. In particular, it is more general than the Jaynes-Cummings model \cite{jcm}, since the Rabi model explicitly includes the counter-rotating terms in the qubit-oscillator interaction \cite{zueco2009qubit}. Hence, it is crucial for describing strong and ultra-strong coupling regimes in which the qubit-oscillator coupling becomes about an order of magnitude smaller than the characteristic frequency of both qubit and  oscillator. These models help to interpret the experiments that have reached the strong coupling regime for various QED setups \cite{wallraff2004strong} and the ultra-strong one in superconducting QED. In the latter case the coupling between artificial atoms and resonators is obtained, for instance, by inductively coupling a flux qubit and an LC oscillator via Josephson junctions \cite{yoshihara2017superconducting}. The ultra-strong coupling regimes of light-matter interaction will keep on expanding at the frontier of quantum optics and quantum physics. Thus, topics related to these regimes will remain a prominent field in the foreseeable future \cite{forn2019ultrastrong}. The other issue to address for a realistic description of such systems is to assess quantitatively the role of dissipation and decoherence \cite{gu2017microwave,perroni2016thermoelectric,nocera2016charge,perroni2021theoretical}. \\
Coupled qubit-resonator systems are currently employed in circuit QED setups since the qubit readout through the Stark dynamical shift of the resonator frequency provides information on the spin component along the quantization axis \cite{gu2017microwave}. This procedure is strictly valid only in the weak coupling limit between qubit and resonator and when interactions with the environment are neglected. Actually, in the weak coupling regime, the closed Rabi model reduces to the closed Jaynes-Cummings model \cite{jcm}, for which it is possible to show that the splitting of the energy eigenvalues depends on qubit state, allowing its readout through the resonator state \cite{zagoskin}. \\
Recently, the readout procedure has been generalized in some experiments \cite{hacohen2016quantum,campagne2016observing,ficheux_tomo} showing that it is possible to follow simultaneously non-commuting spin components of a qubit through resonator observables. In particular, in Ref.~\onlinecite{ficheux_tomo}, the procedure for a full state reconstruction of a superconducting qubit has been improved. These recent experiments have inspired the theoretical work presented in this paper.  However, our theoretical analysis does not explicitly include the effects of the measurement apparatus and hence, even if the physical idea beyond the experiment in Ref.~\onlinecite{ficheux_tomo} is the same, we cannot make a one-to-one correspondence between our work and the experimental realization.
Moreover, after the exposition of the theoretical results, in the concluding section \ref{sec:conclusioni}, we will further comment on the experiments \cite{hacohen2016quantum,campagne2016observing,ficheux_tomo}.\\

In this paper, we make an accurate theoretical analysis of static and dynamics properties of the open quantum Rabi system, from the weak up to the ultra-strong coupling regimes. The oscillator frequency is higher than the qubit one, so that the system is in the anti-adiabatic regime. In the limit of weak internal coupling, this corresponds to the so called dispersive regime\cite{zagoskin}. We use both numerically exact and approximate methods for solving the dynamics with the aim of establishing the validity limits of different methods \cite{breuer2016colloquium,weimer2021simulation}. Firstly, we solve the Heisenberg equations of motion (HEM) in the weak coupling limit between the qubit and the oscillator, in the presence of Ohmic bath. These allow us to derive analytical functional relationships directly linking oscillator and qubit coordinates. Hence, to test the validity of the derived relationships, we perform extensive Matrix Product State (MPS) simulations. We further corroborate our MPS numerical results by solving Lindblad master equations (LME) for the coupled qubit-oscillator system. Only MPS simulations provide correct results in every parameter region, being variationally exact in both the qubit-oscillator coupling and coupling to the environment. Indeed, LME is obtained in the weak coupling limit between the oscillator and the environment and it is valid under the Markovian approximation \cite{zagoskin}. Therefore, in the following, LME results help us assess the non-Markovianity of the solutions obtained with MPS numerics.\\
For what concerns dissipation and decoherence, we analyse the effects of the coupling between the oscillator and the Ohmic bosonic bath. In particular, we show how both the qubit and the oscillator dynamics differ from the free case. Indeed, the systems lose purity in time because of the growing entanglement due to the interplay between the internal and the external couplings. We are interested in regimes of parameters in which these effects are not detrimental for the methods we use to evaluate the mean values of the qubit components through the oscillator dynamics. Moreover, we observe how the presence of the bath helps to start earlier to follow the qubit dynamics, making more immediate the relationships linking qubit and oscillator observables.\\
First, we analyse the more direct relationship between qubit and oscillator. Actually, we make a qubit $z$-component evaluation up to strong coupling regime by making a Fourier transform of the average number of quanta in the oscillator for a qubit in the up state. Then, we extend the procedure to a generic superposition of up and down qubit states including the effects of the bath. Furthermore, we analyse the oscillator dynamics in order to evaluate the qubit state during its evolution. In fact, in addition to the number of bosons, we compute the mean values of the oscillator coordinates. We find that a Bloch vector coordinates evaluation can be achieved by looking at the mean values of the oscillator observables from weak to strong coupling regime if the oscillator is not strongly damped. We quantify the quality of the Bloch vector evaluation by monitoring the qubit fidelity with respect to free limit. We show that, up to the strong coupling regime, we can infer accurately the qubit state from the functional relationships of the Heisenberg equations of motion. Quite surprisingly, results reveal that dissipation and decoherence can facilitate the prediction of qubit dynamics in the parameter regime from weak to strong coupling since they damp time components with the oscillator frequency. Indeed, qubit-oscillator relationships become more direct on the stationary state in comparison with the case of a closed system. \\
Finally, in the ultra-strong coupling regime, the qubit dynamics cannot be separated from the oscillator evolution due to an enhancement of their entanglement. As expected, the Bloch vector
evaluation from the oscillator dynamics becomes
prohibitive in this regime. The Wigner quasi-probability distribution for the oscillator shows that it is in a damped momentum squeezed state, whose squeezing parameter depends on both the internal and the external couplings. Moreover, from the comparison with MPS simulations, we observe how Lindblad equation fails to provide correct results due to the increased role of oscillator-bath interaction. \\
The results presented in this work are not limited to the anti-adiabatic regime. Actually, we will show how a full Bloch vector evaluation can be extended to a parameter regime where the qubit energy scale is of the order of that of the oscillator in the weak to strong coupling regime. \\
The present work can be useful for better understanding the dissipative quantum Rabi model from a fundamental point of view. Furthermore, it can help to address quantum control issues, which are aimed to gain information about system dynamics enabling more powerful performance in computing, sensing, and metrology. \\

The paper is organised as follows: in Sect.~\ref{sec:model_met} we will describe the dissipative quantum Rabi model and the method used to solve the dynamics of this open quantum system; in Sect.~\ref{sec:risultati}, we will discuss our results for three regimes of parameters: weak, strong, and ultra-strong coupling, which are summarized in Fig.~\ref{fig:diag_fase} in the concluding section \ref{sec:conclusioni}. In the App.~\ref{app:closedRabi} we analyse the closed Rabi model which gives indications of the importance of counter-rotating terms; in the App.~\ref{app:Lind} we describe our numerical method to solve the global Lindblad master equation with full secular approximation; in the App.~\ref{app:mps} we present our numerical implementation for MPS simulations; in the last App.~\ref{app:temp} we analyse the robustness of our results with respect to the effect of the temperature.

\section{\label{sec:model_met} Dissipative quantum Rabi model and solution methods}
In this section, we describe the quantum Rabi model, where a harmonic oscillator is coupled to a two-level system (TLS) through a transversal interaction:
\begin{equation}
\label{eq:closedH}
    \mathcal{H}_{closed}=\,\frac{\Delta}{2}\sigma_z+\omega_0 a^{\dagger}a+g\sigma_x (a+a^{\dagger}),
\end{equation}
where $\sigma_z$ and $\sigma_x$ are the canonical Pauli matrices and $\Delta$ is the qubit frequency. In fact, in the $\sigma_z$ eigenbasis it is the gap between the two eigenenergies, while in the $\sigma_x$ eigenbasis it represents the tunneling between the two levels. Then, $a\,(a^{\dagger})$ is the annihilation (creation) operator for the oscillator with frequency $\omega_0$ and $g$ stands for the strength of the qubit-oscillator coupling. The effects of decoherence and dissipation induced by the bosonic environment are taken into account via a linear coupling \textit{à la} Caldeira-Leggett to a collection of $N$ independent bosonic modes at zero temperature:
\begin{align}
    \mathcal{H}_{bath}=&\,\sum_{j=1}^N \left [\omega_j a^{\dagger}_j a_j +\frac{x_0^2}{2} M_j\omega_j^2\right]\\
    &- (a+a^{\dagger})\sum_{j=1}^N\left\| c_j\right\| (a_j+a^{\dagger}_j).
\end{align}
The bosonic modes have frequencies $\omega^2_j=k_j/M_j$, coordinates and momenta given by $x_j$ and $p_j$, respectively; furthermore $x_0$ denotes the position operator of the oscillator with mass $m$ and frequency $\omega_0$: $x_0 = \sqrt{\frac{\hbar}{2m\omega_0}}(a + a^{\dagger})$. Units are such that $\hbar = k_B = 1$. The coupling constants to the bath are $\left\| c_j\right\|=\sqrt{\frac{k_j\omega_j}{4m\omega_0}}$. Moreover, we have neglected the energy shift $\sum_{j=1}^N \omega_j/2$, which does not affect the dynamics. Hence, we can rewrite the Hamiltonian of the system plus the environment, by defining a renormalized oscillator frequency $\bar{\omega}_0=\sqrt{\omega_0^2+\sum_{j=1}^N M_j\omega_j^2/m}$ that encompasses the term $\frac{x_0^2}{2} M_j\omega_j^2$ which ensures that the total energy is bounded from below and the quadratic form is positive definite. It is very natural in superconducting circuits where it leads to the quadratic correction of bosonic modes that ensures that the resonance of the cavity does not change its value, irrespective of the dissipation strength. We also define renormalized coupling strengths $\bar{g}=g\sqrt{\frac{\omega_0}{\bar{\omega}_0}}$, between qubit and oscillator, and $\left\|\bar{g}_j\right\|=\sqrt{\frac{k_j\omega_j}{4m\bar{\omega}_0}}$, between oscillator and each bath bosonic mode. The total Hamiltonian then reads:

\begin{align}
    \label{eq:Hamiltonian}
    \mathcal{H}=\,&\frac{\Delta}{2}\sigma_z+\bar{\omega}_0 b^{\dagger}b+\bar{g}\sigma_x (b+b^{\dagger})\\\nonumber
    +\,&\sum_{j=1}^N \omega_j a^{\dagger}_j a_j
- (b+b^{\dagger})\sum_{j=1}^N \left[\left\|\bar{g}_j\right\|(a_j+a^{\dagger}_j)\right], 
\end{align}

where $b\,(b^{\dagger})$ is the annihilation (creation) operator for the renormalized oscillator with frequency $\bar{\omega}_0$ and coordinates $x=\sqrt{\frac{1}{2m\bar{\omega}_0}}(b + b^{\dagger})$ and $p=i\sqrt{\frac{m\bar{\omega}_0}{2}}(b^{\dagger}-b)$. The bath is then represented by an Ohmic spectral density: $J(\omega) = \sum_{j=1}^N\left\|\bar{g}_j\right\|^2\delta(\omega-\omega_j)=\frac{\alpha}{2}\omega\Theta(\omega_c-\omega)$, $\omega_c$ is the cutoff frequency, and $\theta(x)$ Heaviside function. Here the dimensionless parameter $\alpha$ measures the strength of the oscillator-bath coupling (see Fig.~\ref{fig:model2}). \\
This model can also be mapped \cite{mapzueco} in such a way that, by including the oscillator as a further bosonic mode of the bath, the qubit is coupled to the N+1 bath modes as follows:
\begin{equation}
    \label{eq:mappedH}
    \mathcal{H}_{map}=\frac{\Delta}{2}\sigma_z+\sum_{l=1}^{N+1}\hat{\omega}_l b^{\dagger}_l b_l+\sigma_x \sum_{l=1}^{N+1}\left[\left\|\beta_l\right\|(b_l+b^{\dagger}_l)\right],
\end{equation}
where now the N+1 bosonic modes are defined by the operators $b_l$ and $b^{\dagger}_l$ and have frequencies $\hat{\omega}_l$. The couplings between the qubit and each bosonic mode $\left\|\beta_l\right\|$ are defined such that we can describe the bath in terms of an effective spectral density:
\begin{align}
    J_{eff}(\omega)=&\sum_{l=1}^{N+1}\left\|\beta_l\right\|^2\delta(\omega-\hat{\omega}_l)\\\nonumber
    &\xrightarrow[N\rightarrow\infty]{}\frac{2g^2\omega_0^2\omega\alpha}{(\omega_0^2-\omega^2)^2+(\omega_0\alpha\pi\omega)^2},
\end{align}
which is Ohmic at low frequencies: $J_{eff}(\omega) \approx \frac{2 g^2 \alpha}{\omega_0^2}\omega$. Therefore, the qubit is coupled to an oscillator bath through an effective constant proportional to $g^2 \alpha/ \omega_0^2$. Indeed, in the anti-adiabatic regime ($\omega_0$ large), when the qubit-oscillator coupling $g$ is small, the qubit is only weakly coupled to its surroundings.\\

\begin{figure}[t]
\centering
\includegraphics[scale=0.4]{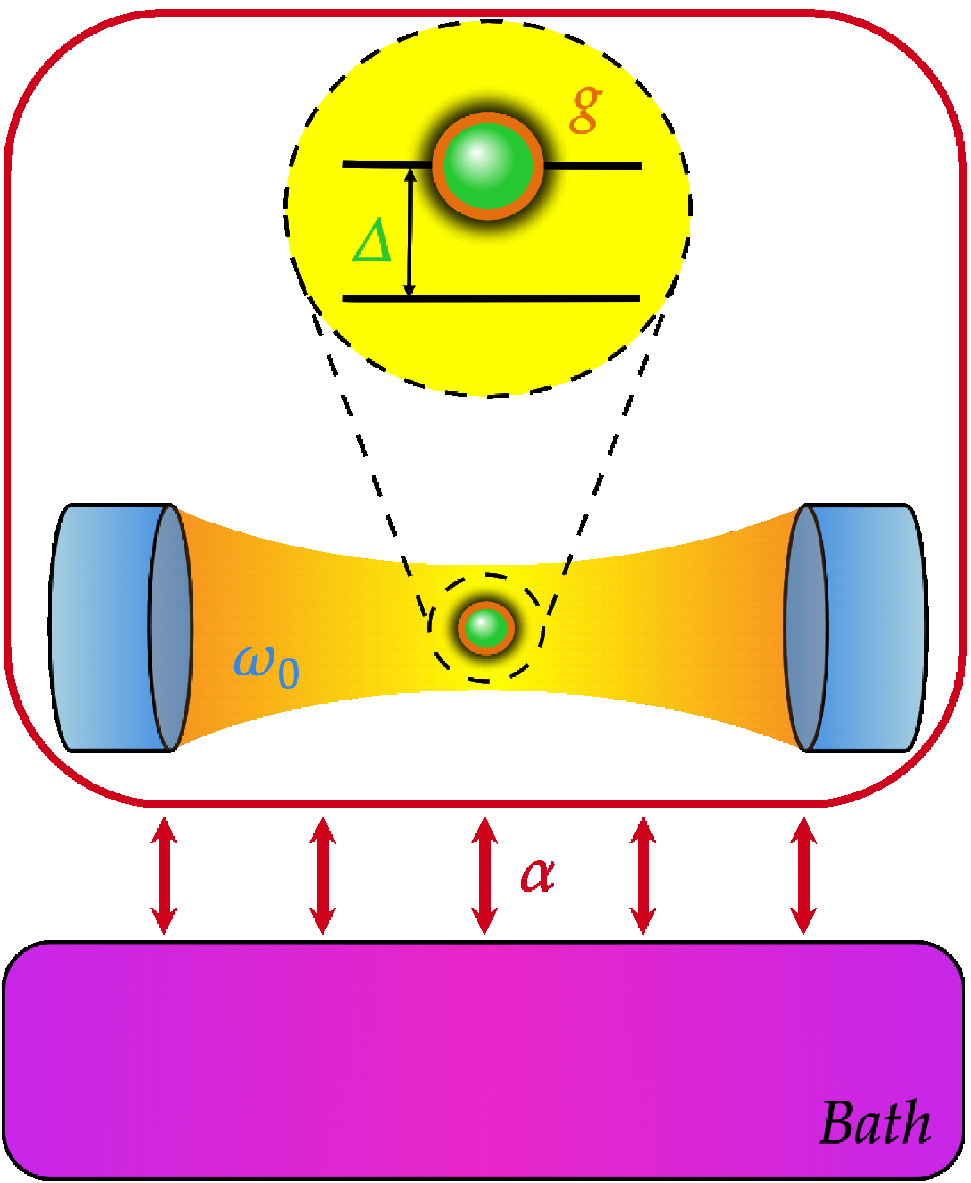}
\caption{\label{fig:model2}Dissipative quantum Rabi model described by the Hamiltonian in Eq.~(\ref{eq:closedH}) for a qubit of frequency $\Delta$ coupled through $g$ to a harmonic oscillator of frequency $\omega_0$, in turn coupled to a bath via a coupling strength $\alpha$}
\end{figure}

In the following subsections we are going to show the approaches useful to describe the time-evolution of the systems of interest (qubit and oscillator) when coupled with the environment through the oscillator. First, we will find the solution of the HEM for position and momentum of the oscillator, perturbative in the coupling strength $g$. Then, we solve the global dynamics of the qubit-oscillator system coupled with the bath using variationally exact MPS simulations and the LME approach. The MPS method allows us to explore the dynamics of the system beyond the perturbative limit in the coupling with the bath. Indeed, as the Hilbert space of the problem grows exponentially with system size, the use of exact diagonalization is prohibitive. \\
In App.~\ref{app:Lind}, we expose the solution of global LME for the composite system qubit plus oscillator. Moreover, in App.~\ref{app:mps} we present the method that we adopt for the dynamic simulations through MPS. \\
In this section, we underline the regimes in which the three methods provide a correct description. In particular, we are interested in the effects of dissipation of the $z$-component evaluation procedure. Moreover, thanks to the MPS simulations, we will be able to verify the functional relationships provided by HEM, useful to unveil the regimes where a full Bloch vector evaluation is possible, which are summarized in the phase diagram in Fig.~\ref{fig:diag_fase}. Actually, MPS allows us to go beyond the other two perturbative methods, correctly analysing strong and ultra-strong coupling regimes, when the effect of the bath is not negligible.

\subsection{Perturbative solution of HEM}
This approach is possible since our bath is made up of bosonic modes linearly coupled to the system made of qubit plus oscillator. Hence, we can exactly eliminate the dynamics of the bath from HEM for the variables pertaining to the reduced system (qubit and oscillator). Starting from the Hamiltonian $\mathcal{H}$ in Eq.~(\ref{eq:Hamiltonian}) and evaluating the HEM $\dot{A}(t)=i\left[\mathcal{H},A(t)\right]$ for a generic observable $A$ belonging to the qubit-oscillator-bath system, we find:\\
\vspace{-10pt}
\begin{equation}
\begin{cases}
\label{eq:bathHEM}
    \dot{x}_j(t)=\frac{p_j(t)}{M_j}\\
    \dot{p}_j(t)=-M_j\omega_j^2 x_j(t)+\lambda_j x(t)\,;
\end{cases}\\
\end{equation}
\noindent
\begin{equation}
\begin{cases}
\label{eq:oscHEM}
    \dot{x}(t)=\frac{p(t)}{m}\\
    \dot{p}(t)=-m\bar{\omega}_0^2 x(t)-\bar{g}\sigma_x(t)\sqrt{2m\bar{\omega}_0}+\sum\limits_{j=1}^N\lambda_j x_j(t).
\end{cases} 
\end{equation}
\noindent
The first system of equations refers to the bath positions and momenta (Eq.~(\ref{eq:bathHEM})), while the second one is for the oscillator observables (Eq.~(\ref{eq:oscHEM})). The coefficients that appear in both the systems are $\lambda_j=-\bar{g}_j\sqrt{2m\bar{\omega}_0}\sqrt{2M_j\omega_j}$. These differential equations are coupled, such that we can express for example $x(t)$ from the first system as a function of the $x_j(t)$ and substitute it in the second one. Therefore, we obtain an equation for $x(t)$ (and similarly for $p(t)$) as follows:
\begin{equation}
    \begin{split}
    \ddot{x}(t)+\bar{\omega}_0^2 x(t)+\bar{g}\sigma_x(t)\sqrt{\frac{2\bar{\omega}_0}{m}}+\frac{d}{dt}\int_0^t ds \gamma(t-s)x(s)\\
    =\frac{1}{m}\sum_{j=1}^N\frac{\lambda_j}{\sqrt{2M_j\omega_j}}\left(a_je^{-i\omega_j t}+a^{\dagger}_j e^{i\omega_j t}\right)=\frac{B(t)}{m}.
    \end{split}     
\end{equation}
It may be viewed as the quantum analogue of a classical stochastic differential equation\cite{breuer}, involving a damping kernel 
\begin{equation}
    \gamma(t-s)=4\bar{\omega}_0\int_0^{\infty}d\omega \frac{J(\omega)}{\omega}\cos\left[\omega\left(t-s\right)\right]
\end{equation} and a force operator $B(t)$ that on average becomes a stochastic force, whose statistics depends on the initial bath states distribution. In our case, when the Ohmic spectral density has an infinite cutoff, $\omega_c \rightarrow \infty$, we get $\gamma(t-s)\rightarrow 4\gamma\delta(t-s)$, where $\gamma=\pi\frac{\alpha}{2}\bar{\omega}_0$ is the oscillator decaying rate. In all the approaches used in this work, we consider the limit where $\omega_c$ is the highest frequency scale. This approximation allows us to simplify the derivative of the integral with the damping kernel to $2\gamma\dot{x}(t)$. Thus, we perform the average over the initial pure separable state of the whole system and find the HEM for the oscillator coordinates:
\begin{align}
    \left\langle\ddot{x}(t)\right\rangle+\bar{\omega}_0^2\left\langle x(t)\right\rangle+2\gamma\left\langle\dot{x}(t)\right\rangle=&-\bar{g}\left\langle\sigma_x(t)\right\rangle\sqrt{\frac{2\bar{\omega}_0}{m}}\\
    \left\langle\ddot{p}(t)\right\rangle+\bar{\omega}_0^2\left\langle p(t)\right\rangle+2\gamma\left\langle\dot{p}(t)\right\rangle=&\bar{g}\left\langle\sigma_y(t)\right\rangle\sqrt{2\bar{\omega}_0 m},
\end{align}
where the third terms are friction forces and the fourth ones are ``external" forces due to the interaction with the qubit. Clearly the solution of these coupled differential equations ($\left\langle \dot{x}(t)\right\rangle=\frac{\left\langle p(t)\right\rangle}{m}$) depends on the time evolution of the mean values of the qubit Pauli matrices. A way to find an analytical solution in the anti-adiabatic regime ($\Delta\ll \bar{\omega}_0$) is to consider the ratio between the qubit-oscillator coupling and the oscillator frequency, $\bar{g}/\bar{\omega}_0$ as a weak perturbation. By neglecting terms greater than or equal to the first order in $\bar{g}/\bar{\omega}_0$ in the oscillator equations, the HEM for the Pauli matrices can be solved at zero order in $\bar{g}/\bar{\omega}_0$, describing the well-known Rabi oscillations for a free evolving qubit. Apart from a decaying homogeneous solution on a transient time of the order of $~\gamma^{-1}$, the solutions for the mean dimensionless oscillator position $\tilde{x}=x\sqrt{2m\bar{\omega}_0}$ and momentum $\tilde{p}=p\sqrt{2/(m\bar{\omega}_0)}$ are the following:
\begin{align}
\label{eq:partsolx}
    \left\langle \tilde{x}(t)\right\rangle\rightarrow&\,-2\bar{\omega}_0 k \left\langle \sigma_x(t) \right\rangle\\
\label{eq:partsolp}
    \left\langle\tilde{p}(t)\right\rangle\rightarrow&\,2\Delta k \left\langle \sigma_y(t) \right\rangle,
\end{align}
where the coefficient $k$ is 
\begin{equation}
k= \frac{\bar{g}(\bar{\omega}_0^2-\Delta^2)}{\Delta^4+\bar{\omega}_0^4+4\gamma^2\Delta^2-2\Delta^2\bar{\omega}_0^2}.
\end{equation}
\\
At the zero-order in the coupling $\bar{g}$, the mean Pauli matrices are those of a free spin 
\begin{align}
\label{eq:sigmax}
    \left\langle \sigma_x(t)\right\rangle=&\, \left\langle\sigma_x(0)\right\rangle \cos(\Delta t)-\left\langle \sigma_y(0)\right\rangle \sin(\Delta t)\\
\label{eq:sigmay}
    \left\langle \sigma_y(t)\right\rangle=&\, \left\langle\sigma_y(0)\right\rangle \cos(\Delta t)+\left\langle \sigma_x(0)\right\rangle \sin(\Delta t)\\
\label{eq:sigmaz}
    \left\langle \sigma_z(t)\right\rangle=&\, \left\langle\sigma_z(0)\right\rangle.
\end{align}
In Sect.~\ref{sec:risultati}, we will show that, in the weak coupling regime, this treatment of the mean values is sufficient to get accurate results.\\
In the next section, thanks to the MPS simulations, we will show that
Eqs.~(\ref{eq:partsolx}-\ref{eq:partsolp}) are valid up to the strong coupling regime. Moreover, we will show that the main effects of the non-zero $\bar{g}$ are to introduce a renormalized frequency ($\Delta_x$ and $\Delta_y$ for $\left\langle \sigma_x(t)\right\rangle$ and $\left\langle \sigma_y(t)\right\rangle$ respectively) and a decay rate ($\kappa_x$ and $\kappa_y$ for $\left\langle \sigma_x(t)\right\rangle$ and $\left\langle \sigma_y(t)\right\rangle$ respectively) in Eqs.~(\ref{eq:sigmax}-\ref{eq:sigmay}):
\begin{align}
\label{eq:sigmarenx}
    \left\langle \sigma_x(t)\right\rangle= \left(\left\langle\sigma_x(0)\right\rangle \cos(\Delta_x t)-\left\langle \sigma_y(0)\right\rangle \sin(\Delta_x t)\right)e^{-\kappa_x t}\\
\label{eq:sigmareny}
    \left\langle \sigma_y(t)\right\rangle= \left(\left\langle\sigma_y(0)\right\rangle \cos(\Delta_y t)+\left\langle \sigma_x(0)\right\rangle \sin(\Delta_y t)\right)e^{-\kappa_y t},
\end{align}
while for the $z$-component, a good fit is obtained by using a linear regression with slope $a_z$ ($<0$) and intercept $b_z$:
\begin{equation}
    \label{eq:sigmarenz}
    \left\langle \sigma_z(t)\right\rangle=\, a_z t+b_z.
\end{equation}
The intercept is the initial value of the $z$-component and this regression is valid up to our final time of numeric simulations, that is $t\approx5T$, where $T$ is the qubit period. We note how the effect of the bath on the qubit is to reduce its purity in time so that the Bloch vector is no more on the Bloch sphere surface, but its point is inside the sphere. Nevertheless, in the regimes of parameters in which we can do a reliable Bloch vector evaluation, the qubit is still close to its free evolution and hence next to the Bloch surface.
We will show that the relationships above fit very well the results obtained with MPS approach in Sect.~\ref{sec:risultati}.\\
There remains the problem of evaluating the $z$-component of the qubit trajectory in the Bloch sphere. An approximate time evolution can be written starting from the average number of quanta in the oscillator in terms of its coordinates and their standard deviations $\delta\tilde{x}$ and $\delta\tilde{p}$. By using Eqs.~(\ref{eq:partsolx}-\ref{eq:partsolp}), one obtains:
\begin{eqnarray}
    \left\langle \tilde{n}(t)\right\rangle=&\bar{\omega}_0^2 k^2(1-\left\langle\sigma_z(t)\right\rangle^2)+k^2(\bar{\omega}_0^2-\Delta^2)\left\langle\sigma_y(t)\right\rangle^2\nonumber\\
    +&\hspace{-70pt}\frac{1}{4}\left((\delta \tilde{x}(t))^2+(\delta \tilde{p}(t))^2\right)-\frac{1}{2}.
    \label{eq:NHEM}
\end{eqnarray}
Since $\left\langle\sigma_y(t)\right\rangle$ can be derived from previous Eqs.~(\ref{eq:partsolx}-\ref{eq:partsolp}), we get access to $\left\langle\sigma_z(t)\right\rangle$ through the oscillator observables over time. In fact, we do not need to infer its sign, but the squared value is sufficient. That is because if $g$ can still be treated perturbatively, $\left\langle\sigma_z\right\rangle$ decreases in time linearly with a small slope in absolute value, oscillating with small oscillation amplitude, not changing sign. Moreover, starting from a qubit state on the equator of the Bloch sphere, $\left\langle\sigma_z\right\rangle$ remains zero up to the strong coupling regime for all times, if the coupling to the bath is small. In particular, as found in the next section, the third term in Eq.~(\ref{eq:NHEM}) is only a small correction, but it oscillates in time with the frequencies of the coupled system that allow us to read the qubit state. \\
In the case of a qubit state initially up along $z$ direction, Eqs.~(\ref{eq:sigmax}-\ref{eq:sigmay}), but also Eqs.~(\ref{eq:sigmarenx}-\ref{eq:sigmareny}), vanish, therefore, Eq.~(\ref{eq:NHEM}) provides a direct link between the average number of quanta in the oscillator and the $z$-component of the spin. This is our procedure to evaluate the $z$-component of the qubit in the Bloch sphere in time. On the other hand, if the qubit is initially in a linear combination of up and down states along $z$ direction, Eqs.~(\ref{eq:partsolx}-\ref{eq:partsolp}) are different from zero, hence the mean number of bosons in Eq.~(\ref{eq:NHEM}) displays a more complex time behaviour. \\
One of the aims of this paper is to characterise the effects of the bath on this procedure to read the state of a qubit initially in a linear combination of up and down states. Ultimately, the approximate but analytical solutions in Eqs.~(\ref{eq:partsolx}-\ref{eq:partsolp}-\ref{eq:NHEM}) are the functional relationships linking qubit and oscillator observables which will guide us in the remaining part of the paper. These equations explicitly show how it is possible with proper tuning to follow the qubit state through the oscillator time evolution. We will follow essentially this procedure in the next section: we determine oscillator observables from independent MPS simulations, then we will use the functional HEM Eqs.~(\ref{eq:partsolx}-\ref{eq:partsolp}-\ref{eq:NHEM}) to fit the MPS data for the oscillator observables by using MPS results for the spin components. In particular, the parameters of Eqs.~(\ref{eq:sigmarenx}-\ref{eq:sigmareny}) are obtained from MPS simulations. In the next section, we will call this procedure as $HEM+fit$. We will find that, up to the strong coupling regime, the agreement is excellent, implying that, by inverting Eqs.~(\ref{eq:partsolx}-\ref{eq:partsolp}-\ref{eq:NHEM}), the oscillator observables provide access to all three spin components, therefore to the Bloch vector evaluation.

\subsubsection{Some remarks on the perturbative analytical solution} \label{subsec:deco_squeezed}
Let us have a closer look to the solution of the HEM derived above. We stress two points which will be relevant in the next sections:
\begin{itemize}
    \item[1)] dissipation and decoherence induced by the bath on the oscillator are not detrimental to the Bloch vector evaluation;
    \item[2)] in the anti-adiabatic regime, with increasing oscillator-qubit coupling, the oscillator is in a well approximated squeezed state, depending on the qubit state, with a time-dependent squeezing parameter $r(\alpha,\bar{g})$.
\end{itemize}
The first consideration is due to the fact that the homogeneous solution of the HME is proportional to the factor $e^{-\gamma t}$, hence, for $\gamma\rightarrow\infty$ it vanishes. We expect that as the coupling to the bath increases, the Bloch vector evaluation can start earlier, on a time scale of the order of $\tau\approx3/\gamma$.\\
For the second point, it is useful to consider a standard canonical transformation for the oscillator at time $t$ such that the new shifted annihilation operator is $c=b+\frac{\bar{g}}{\bar{\omega}_0}\sigma_x$ and therefore the new oscillator Hamiltonian is $\mathcal{H}_o=\bar{\omega}_0 c^{\dagger}c-\frac{\bar{g}^2}{\bar{\omega}_0}$. We obtain just the Hamiltonian of a shifted harmonic oscillator whose vacuum state is the coherent state centred at $-\frac{\bar{g}}{\bar{\omega}_0}\sigma_x$. Within the anti-adiabatic regime, the qubit is much slower than the oscillator, therefore it can be followed by the oscillator in a coherent state moving according to the $\left\langle\sigma_x\right\rangle$ of the qubit. Indeed, the HEM for qubit and oscillator are coupled so that $\sigma_x$ can be written as a term to the zero-order in $\bar{g}$ plus a term which depends on $ \bar{g}^2\tilde{x}$. As a consequence, a term to the third order in $\bar{g}$ proportional to $\tilde{x}^2$ appears in the oscillator Hamiltonian, indicative of a squeezed state, with squeezing parameter $r(\alpha,\bar{g})$. In particular, being the coupling to the qubit in the oscillator position, we expect a momentum squeezed state.

\section{\label{sec:risultati}Comparison of different regimes}
In this section we showcase our results by applying the HEM method previously introduced and the LME and the MPS ones described in detail in App.~\ref{app:Lind} and \ref{app:mps} respectively to the dynamics of our model, in the anti-adiabatic regime ($\Delta\ll \bar{\omega}_0$), from the weak to the ultra-strong coupling regime. Specifically, up to the strong coupling regime, for weak to intermediate couplings of the oscillator to the bath, we find that the oscillator allows the $z$-component evaluation of the qubit (up or down initial qubit state). Furthermore, through the time evolution of oscillator observables, starting from a generic qubit pure initial state, we evaluate the qubit state by computing the qubit fidelity in time with respect to its free evolution (see Fig.~\ref{fig:diag_fase} for a summary of the results of the results). For parameters from the strong to the ultra-strong coupling regime, the interactions (between them and with the bath) are so important that the qubit behaviour differs from the free one as a result of the oscillator back-action upon the qubit. However, our analysis allows us to go beyond the Lindblad approximation, observing non-Markovian features of the dynamics, even through the Wigner quasi-probability distribution for the oscillator state.\\
This section is organised as follows. In Subsect.~\ref{subsec:WC} we will show how the Bloch vector evaluation of the qubit works in the weak coupling regime for the three techniques (HEM, LME, MPS); we evaluate the qubit fidelity, and clarify in what sense dissipation and decoherence are useful for the Bloch vector evaluation. In Subsect.~\ref{subsec:SC} we will identify the range of bath-oscillator couplings for which a $z$-component evaluation is feasible in the strong coupling regime, for both the up state and the generic one; moreover, we will continue the analysis for the Bloch vector evaluation. In Subsect.~\ref{subsec:USC}, finally, we will discuss the features of the dynamics in the ultra-strong coupling regime, where entanglement and non-Markovianity are non-negligible and neither qubit $z$-component evaluation nor Bloch vector one are achievable.

\begin{figure*}[t]
\centering
\includegraphics[scale=0.28]{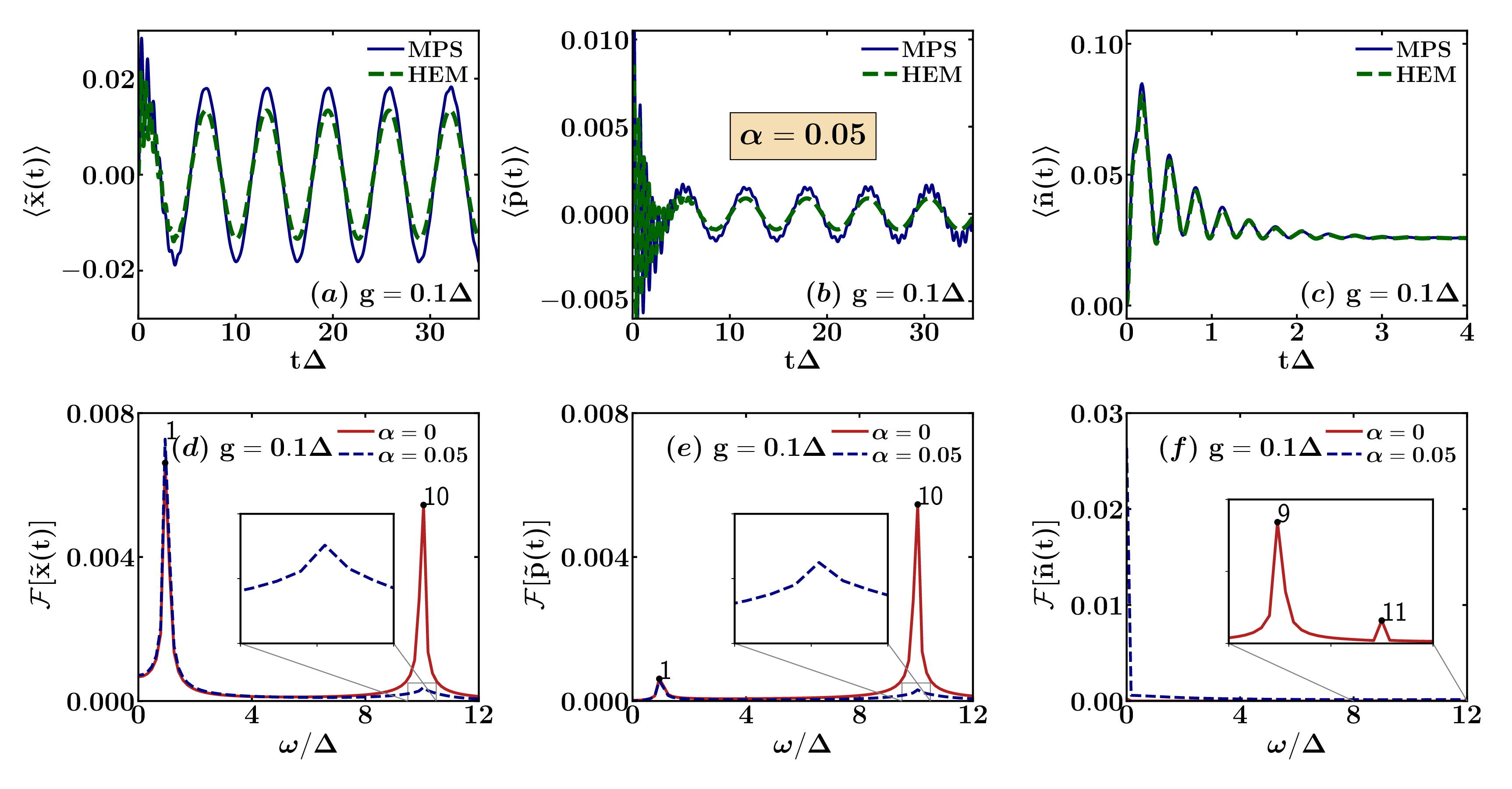}
\caption{\label{fig:xpn_wc} Oscillator position in $(a)$, oscillator momentum in $(b)$ and oscillator number in $(c)$ as a function of time (in units of $1/\Delta$) for two different methods: MPS simulations (blue solid line) and the HEM approach (green dashed line) (particular solutions in Eqs.~(\ref{eq:partsolx}-\ref{eq:partsolp})). FFT of oscillator position in $(d)$, oscillator momentum in $(e)$ and oscillator number in $(f)$ as a function of the frequency (in units of $\Delta$) for the closed case at $\alpha=0$ (red solid line) and the open one $\alpha=0.05$ (blue dashed line) computed at the same parameters through MPS method. Parameters: $\omega_0=10\Delta,g=0.1\Delta,\alpha=0.05,\omega_c=50 \Delta$; for MPS simulations, $N_o=10,N=500,N_{ph}=3,D_{\rm{max}}=20$ }
\end{figure*} 

\subsection{\label{subsec:WC}Weak coupling regime}
We set the parameters such that the qubit frequency $\Delta$ is taken as unity, hence, for $\omega_0=10\Delta$, $g=0.1\Delta$ and $\alpha=0.05$, $g/\Delta\leq\Gamma/\Delta\,\bigcap\,\Gamma/\Delta\leq0.1\omega_0/\Delta$, with $\Gamma=\alpha\omega_0\pi/2$ the bare oscillator decay rate, involving that the quantum Rabi system is in the weak coupling regime \cite{gu2017microwave} (Fig.~\ref{fig:diag_fase}). We remark that, in the limit of small $\alpha$, the agreement between HME, LME and MPS results is excellent. \\
%\newpage
Before analysing the dissipative case, we introduce some features of the closed one, useful to understand what happens when we turn on the coupling to the bath. In App.~\ref{app:closedRabi} we perturbatively solve the closed Rabi model, taking into account only the states with zero or one boson in the oscillator. In this way, we observe that for small values of $g$ the mean oscillator number in time oscillates with the same frequencies of the magnetization and in fact the Fourier transform shows peaks at the frequencies $\omega_0\pm\Delta$, depending on the qubit state. Instead, the $x$ and $y$ components of the Bloch vector have a time-dependent behaviour that is very similar to the one of the free case (Rabi oscillations at frequency $\Delta$). The position and momentum of the oscillator oscillate at frequencies $\omega_0$ and $\Delta$. Hence, we evaluate the $z$-component of the qubit from the oscillator number and we can follow the position and momentum to evaluate the other components at each time.

In this subsection, we set the coupling $\alpha=0.05$, an intermediate value of the oscillator-bath coupling, since it allows to understand the general method used in the next subsections.\\

We present our implementation of Bloch vector evaluation, by following the oscillator state. Starting from the state $\ket{0}\otimes\ket{\nearrow}$, with $\ket{\nearrow}= (\cos(\theta/2)\ket{\uparrow}+\sin(\theta/2)\,e^{-i\phi}\ket{\downarrow})$, where $\theta=0.3\pi$ and $\phi=1.2\pi$, and an empty bath, we use MPS and the complete solutions of HEM (only particular solutions in Eqs.~(\ref{eq:partsolx}-\ref{eq:partsolp})). As shown in Fig.~\ref{fig:xpn_wc}, we find a good agreement between the two approaches. The oscillator starts following the qubit after $\tau\approx 3/\gamma\approx3.12/\Delta$. The agreement is good for the oscillator coordinates, plotted in the panels (a) and (b) of Fig.~\ref{fig:xpn_wc}. There is a small difference in the amplitude of the oscillations which is not quite well reproduced by the HEM approach, because it is proportional to the qubit-oscillator coupling constant. For the number, instead, panel (c) of Fig.~\ref{fig:xpn_wc} shows an excellent agreement. We notice how the number becomes constant, only shifted from $0$ to $0.0257$, after a rapid oscillation. In the panels $(d)$, $(e)$ and $(f)$ we show respectively the Fast Fourier Transform (FFT) of the oscillator observables above. These plots put in evidence the effect of dissipation that is the fallout of the oscillation in the time evolution of oscillator number and of that at oscillator frequency in the time evolution of position and momentum (see App.~\ref{app:closedRabi}). For this reason, we can evaluate the $z$-component of the qubit from the residual oscillations in the FFT of the oscillator number, but only for small values of $\alpha$ as underlined in the following sections. Alternatively, we can make a Bloch vector evaluation by following the oscillator position and momentum in time after the time transient. 
%\onecolumngrid
%\begin{center}

%\end{center}
%\newpage
%\twocolumngrid
Both LME and MPS show the possibility to follow the qubit and are consistent with each other. They also confirm that the HEM solution can be improved by using the fit for the mean values of spin observables given in Eqs.~(\ref{eq:sigmarenx}-\ref{eq:sigmareny}). In fact, in panel $(a)$ of Fig.~\ref{fig:xF_wc}, we notice this agreement for the time evolution of the oscillator position. For LME and HEM there is a difference with MPS in the amplitude of the oscillations due to a more accurate treatment of the coupling.

\begin{figure}[b]
\centering
\includegraphics[scale=0.215]{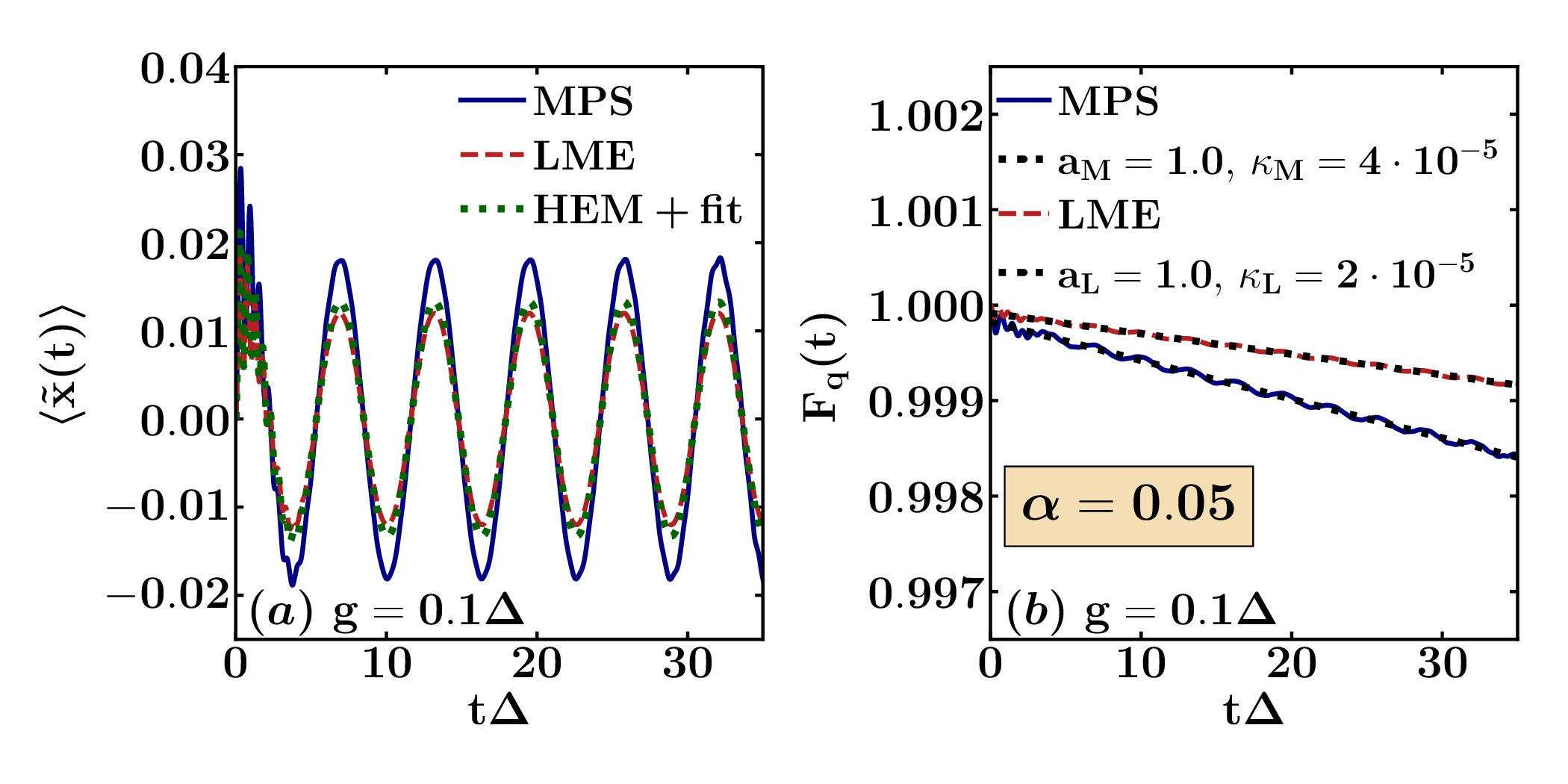}
\caption{\label{fig:xF_wc} Oscillator position in $(a)$ and qubit fidelity in $(b)$ as a function of time (in units of $1/\Delta$). In panel $(a)$ the oscillator position is computed with three methods: MPS simulations (blue solid line), LME solution (red dashed line) and HEM approach (green dotted line), improved by the fit given in Eqs.~(\ref{eq:sigmarenx}-\ref{eq:sigmareny}). Parameters: $\omega_0=10\Delta,g=0.1\Delta,\alpha=0.05,\omega_c=50 \Delta$; for MPS simulations, $N_o=10,N=500,N_{ph}=3,D_{\rm{max}}=20$}
\end{figure}
%\newpage
We assess the quality of the Bloch vector evaluation by evaluating the qubit fidelity $F_q(t)$ with respect to its free evolution, when decoupled from the oscillator:
 \begin{equation}
     \rho^{free}(t)=\frac{1}{2}\left(\mathbf{I}+\left\langle\vec{\sigma}^{free}(t)\right\rangle\cdot\vec{\sigma}^{free}\right),
 \end{equation}
 where the free Bloch vector describes the Rabi oscillations and it is a pure state. Thus, the qubit fidelity reads:
\begin{align}
    F_q(t)=&\left(\Tr\sqrt{\sqrt{\rho^{free}(t)}\rho(t)\sqrt{\rho^{free}(t)}}\right)^2\\
    =&\bra{\psi_{\rho^{free}(t)}}\rho(t)\ket{\psi_{\rho^{free}(t)}}\\
    =&\frac{1}{2}\left(1+\left\langle\vec{\sigma}^{free}(t)\right\rangle\cdot\left\langle\vec{\sigma}(t)\right\rangle\right).
\end{align}
A fidelity above the 90 percent is assumed reliable. Indeed, in this coupling regime, the Bloch vector evaluation is robust, as shown in panel $(b)$ of Fig.~\ref{fig:xF_wc}. This panel also shows a comparison between MPS and Lindblad methods by following a linear fit for both: $F_q(t)\approx a_n-\kappa_n t$, where $n=M$ for fitting of MPS curve, $n=L$ for fitting of LME plot. The Lindblad average fidelity decreases with a slope smaller in absolute value than that computed with MPS. Anyway, both MPS and Lindblad solutions give a fidelity close to $100\%$ up to $t\approx5T$, where $T$ is the qubit period (see Fig.~\ref{fig:diag_fase} and Table~\ref{tab:fid} in the concluding sect. \ref{sec:conclusioni}).\\
\noindent
\begin{figure}[t]
\centering
\includegraphics[scale=0.4]{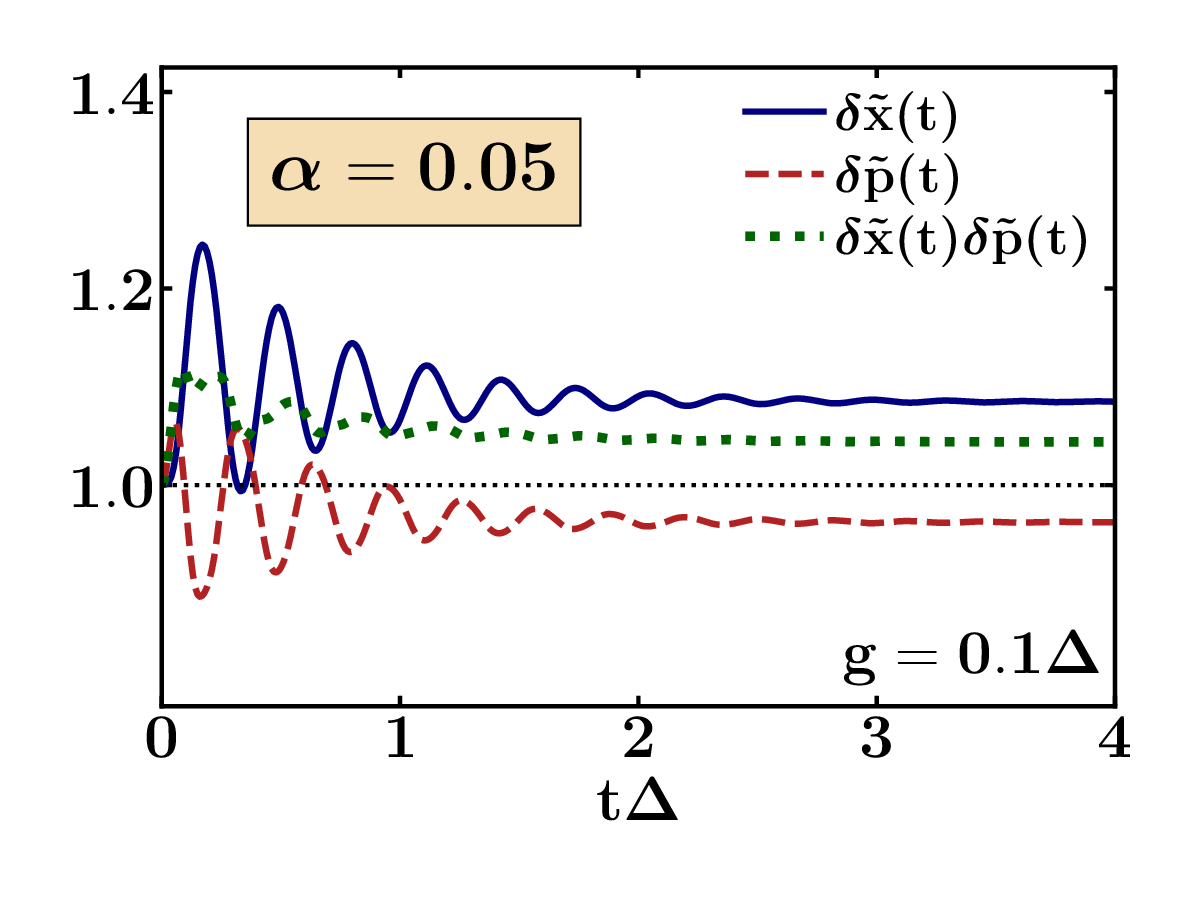}
\caption{\label{fig:indet_wc} Oscillator position uncertainty (blue solid line), momentum one (red dashed line) and uncertainty product (green dotted line) as a function of time (in units of $1/\Delta$) from MPS simulations. Parameters: $\omega_0=10\Delta,g=0.1\Delta,\alpha=0.05,\omega_c=50 \Delta$; for MPS simulations, $N_o=10,N=500,N_{ph}=3,D_{\rm{max}}=20$ }
\end{figure}

With increasing $\alpha$, as expected, the HEM homogeneous solution effectively drops earlier since the transient $\tau\approx 3/\gamma$ gets reduced up to very strong couplings, when the interactions break the qubit-oscillator coherence. Furthermore, starting from the vacuum state, the oscillator evolves in a momentum squeezed state, not so different from the vacuum, due to the small value of $\bar{g}$, as underlined in \ref{subsec:deco_squeezed}. In particular, the figure Fig.~\ref{fig:indet_wc} shows that the position uncertainty $\delta\tilde{x}(t)>1$, the momentum one is instead $\delta\tilde{p}(t)<1$, indication of a momentum squeezed state, so that the product $\delta\tilde{x}(t)\delta\tilde{p}(t)\geq1$, as ruled by the uncertainty principle. 

\subsection{\label{subsec:SC}Strong coupling regime - non-perturbative Bloch vector evaluation}
We choose again the qubit frequency $\Delta$ as unity and $\omega_0=10\Delta$, by tuning $g$ and $\alpha\in[0.01,0.1]$ such that $g/\Delta\geq\Gamma/\Delta\,\bigcap\,g/\Delta\leq0.1\omega_0/\Delta\,\bigcap\,\Gamma/\Delta\leq0.1\omega_0/\Delta$ or only $g/\Delta\leq0.1\omega_0/\Delta$ if $\Gamma/\Delta\geq0.1\omega_0/\Delta$ (see Fig.~\ref{fig:diag_fase}). Hence, the closed quantum Rabi system is in the strong coupling regime. \\
In order to determine the excitation energies of the closed system we directly apply the stationary perturbation theory for small $g$ on the Hamiltonian in Eq.~(\ref{eq:closedH}). The unperturbed Hamiltonian $\mathcal{H}^0=\mathcal{H}_q+\mathcal{H}_o$ reads:
\begin{equation}
    \mathcal{H}^0\ket{n,\pm}_{0}=\left(\pm\frac{\Delta}{2}+\omega_0n\right)\ket{n,\pm}_{0}=E_{n,\pm}^{0}\ket{n,\pm}_{0},
\end{equation}
so that the eigenvalues up to the second order in $g$ are the following:
\begin{equation}
    E_{n,\pm}\approx\pm\frac{\Delta}{2}+\omega_0n\pm g^2\left(\frac{n+1}{\Delta\mp\omega_0}+\frac{n}{\Delta\pm\omega_0}\right).
\end{equation}
Hence the excitation energies can be easily evaluated as $E_{u(d)}=E_{1,-(+)}-E_{0,+(-)}$:
\begin{align}\label{eq:shift2ord}
    E_{u}&=-\Delta+\omega_0-2g^2\left(\frac{1}{\Delta+\omega_0}+\frac{1}{\Delta-\omega_0}\right)\\
    E_{d}&=\Delta+\omega_0+2g^2\left(\frac{1}{\Delta-\omega_0}+\frac{1}{\Delta+\omega_0}\right),
\end{align}
where $E_{u}$ is the energy corresponding to the up state of the qubit, smaller than the oscillator energy, and $E_{d}$ that of the down one, larger than the oscillator energy. Hence, the peaks of the Fourier transform of the number of bosons are located at the frequencies corresponding to the excitation energies $\omega_{u\,(d)}=\mp \Delta + \omega_0\mp2g^2\left(\frac{1}{\Delta\pm\omega_0}+\frac{1}{\Delta\mp\omega_0}\right)$, where the upper sign is for the up state and the lower one for the down state.\\

\begin{figure}[h]
\centering
\includegraphics[scale=0.26]{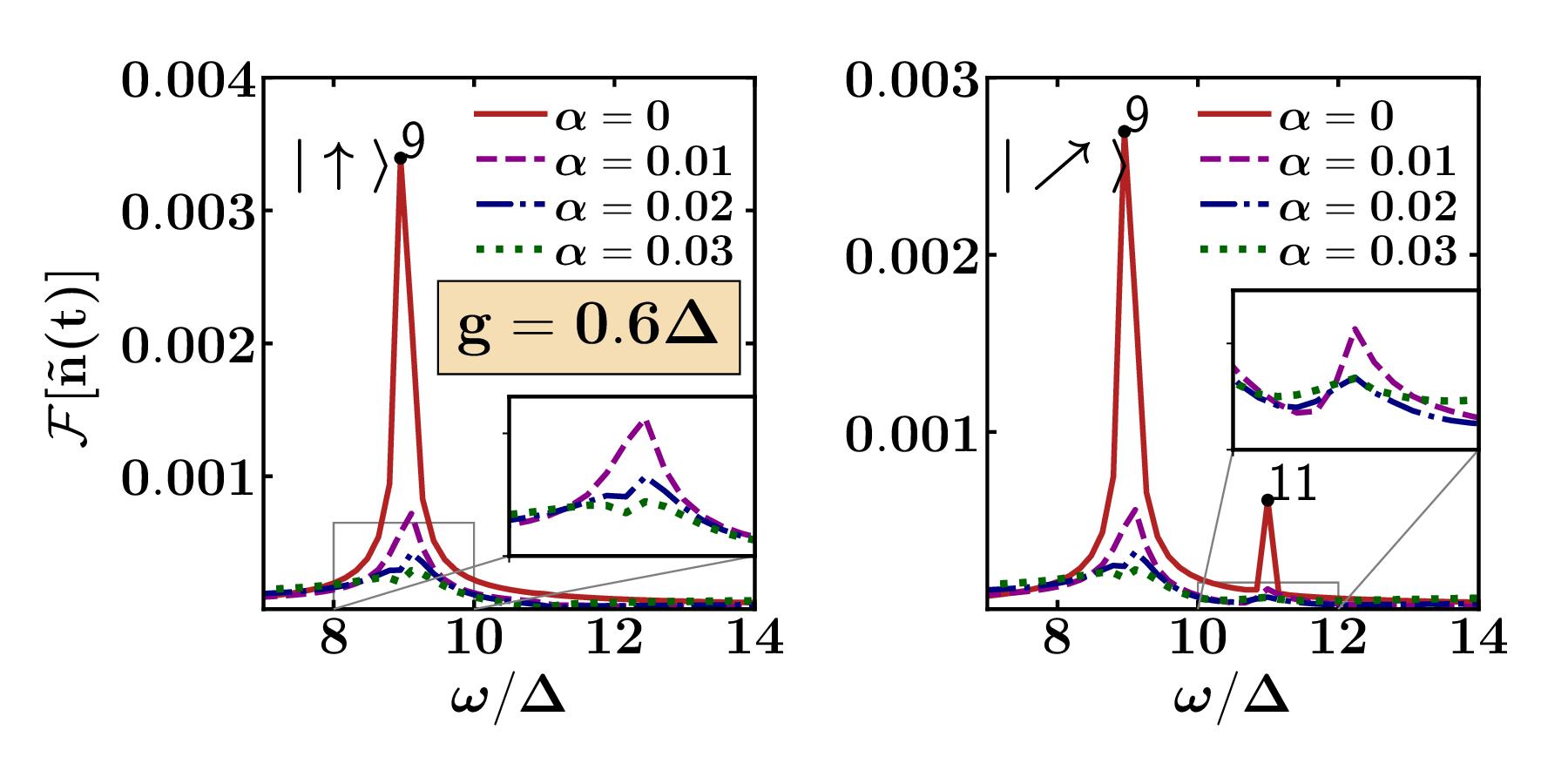}
\caption{\label{fig:FFT_sc} FFT of the mean oscillator number $\tilde{n}$ computed at $g=0.6\Delta$ as a function of frequency (in units of $\Delta$) through MPS simulations for the closed system and for three values of $\alpha$ starting from the two initial states: ($\ket{\uparrow}$ (left panel), $\ket{\nearrow}$ (right panel)). Parameters: $\omega_0=10\Delta,g=0.6\Delta,\alpha\in[0.00,0.03],\omega_c=50 \Delta$; for MPS simulations, $N_o=20,N=500,N_{ph}=3,D_{\rm{max}}=20$}
\end{figure}

We first verify if the $z$-component evaluation through the dissipative oscillator works also in this coupling regime. In particular, we perform the same analysis for a qubit in an up state ($\ket{0}\otimes\ket{\uparrow}$) or in a generic state ($\ket{0}\otimes \ket{\nearrow}$), and an oscillator with zero quanta of energy in contact with the bath at zero temperature. In order to locate the excitation energies of the open system and read the qubit state from the oscillator, we analyse the peaks of the FFT of the mean oscillator number.\\
As expected, we observe in Fig.~\ref{fig:FFT_sc} only a peak for the up state, with frequency lower than the oscillator one, and two peaks for the generic qubit state due to the presence of up (left peak) and down (right peak) contributions. With increasing the oscillator-bath coupling $\alpha$, we find that the peaks become less resolved making the qubit $z$-component evaluation more difficult. In both panels of Fig.~\ref{fig:FFT_sc}, there is a critical value $\alpha_c$ of $\alpha$, about $0.03$, above which the $z$-component evaluation is not feasible. We see how the FFT below the critical value for the generic pure initial state has memory of the down contribution. We also note that the amplitudes reduce by one order of magnitude, when passing from the closed case to the open one for $\alpha=0.01$.\\

\begin{figure}[b]
\centering
\includegraphics[scale=0.3]{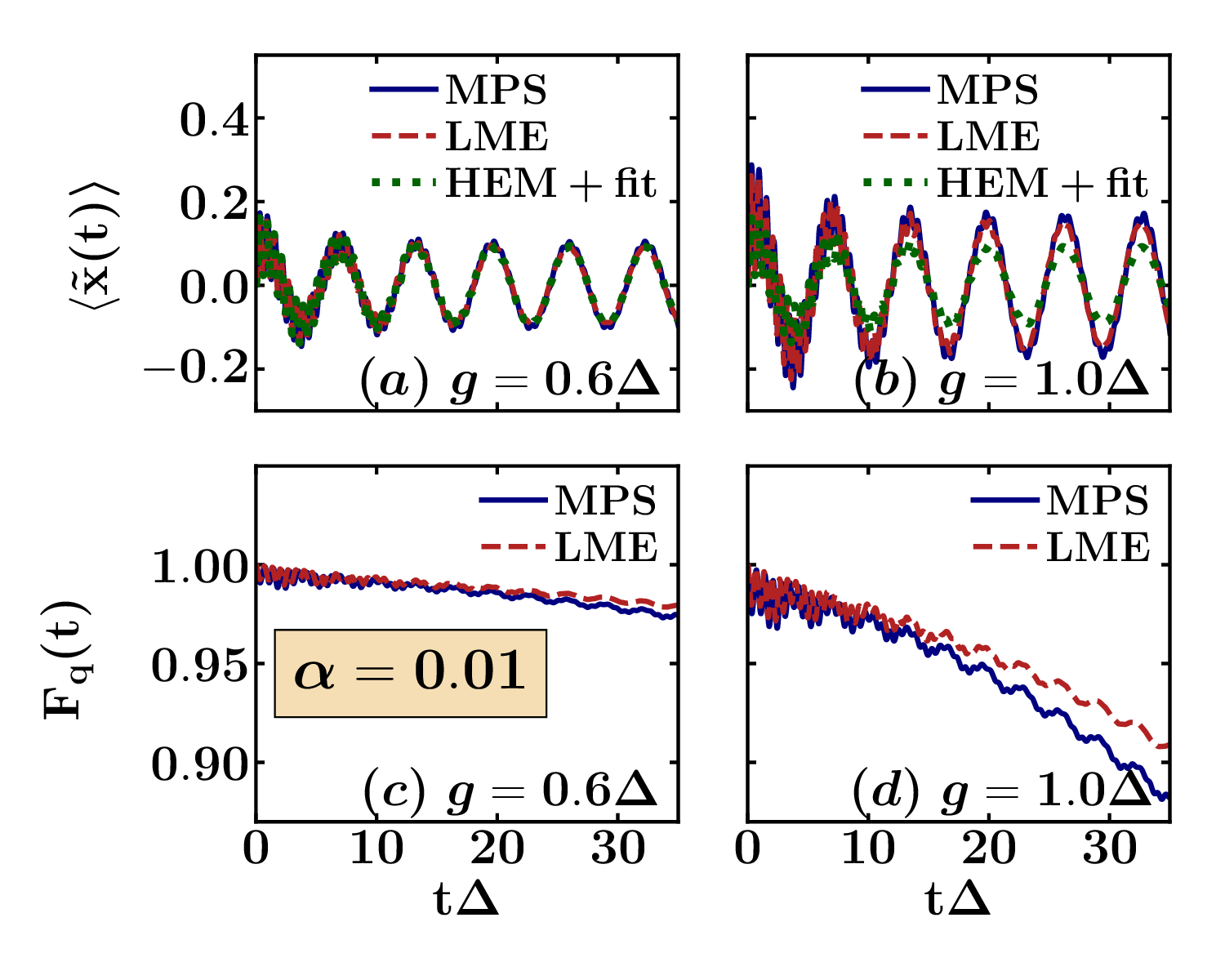}
\caption{\label{fig:xF01_sc} Oscillator position (in panels $(a)$ and $(b)$) and qubit fidelity (in panels $(c)$ and $(d)$) as a function of time (in units of $1/\Delta$) for $g=0.6\Delta$ and $g=1.0\Delta$ at $\alpha=0.01$ using different methods: MPS simulations (blue solid line), the LME solution (red dashed line) and the HEM approach (green dotted line), improved by the fitting with spin observables obtained by MPS simulations as in Eqs.~(\ref{eq:sigmarenx}-\ref{eq:sigmareny}). Parameters: $\Delta=1,\omega_0=10\Delta,g=\{0.6\Delta,1.0\Delta\},\alpha=0.01,\omega_c=50;$ for MPS simulations $N_o=20,N=500,N_{ph}=3,D_{\rm{max}}=20$}
\end{figure}

We then pursue the full Bloch vector evaluation choosing the pure initial state as $\ket{0}\otimes \ket{\nearrow}$ analysing in detail the system behaviour for $g=0.6\Delta$ (the same as in Fig.~\ref{fig:FFT_sc}) and $g=1.0\Delta$, with $\alpha\in\{0.01,0.1\}$. For $\alpha=0.01$, results from LME and MPS are still in agreement and the full state evaluation is possible (see Fig.~\ref{fig:xF01_sc}). \\
%\newpage
At $g=0.6\Delta$, the three approaches give the correct description, because the regime is still perturbative in $g$ and the effect of the bath negligible. At $g=1.0\Delta$, there is a difference in the amplitude of the oscillations of the oscillator position obtained from HEM. Indeed, this solution is strictly valid only for small couplings $g$, unable to significantly influence the qubit dynamics. For the fidelity ($(c)$ and $(d)$ in Fig.~\ref{fig:xF01_sc}) we compare MPS and Lindblad methods. For $g=0.6\Delta$, they overlap almost completely with each other, always above the $95\%$ for $5$ qubit periods. This robust fidelity together with the two peaks of the FFT in Fig.~\ref{fig:FFT_sc} shows how it is possible to still use the HEM approach for oscillator coordinates and energy to evaluate the qubit state in time. Instead, for $g=1.0\Delta$ the system ends up in the limiting case when the Bloch vector evaluation is feasible for both LME and MPS ($F_q(5T\Delta)\approx90\%$). Therefore, it is possible through the oscillator to read the qubit state which is still close to the free behaviour. \\

\begin{figure}[h]
\centering
\includegraphics[scale=0.31]{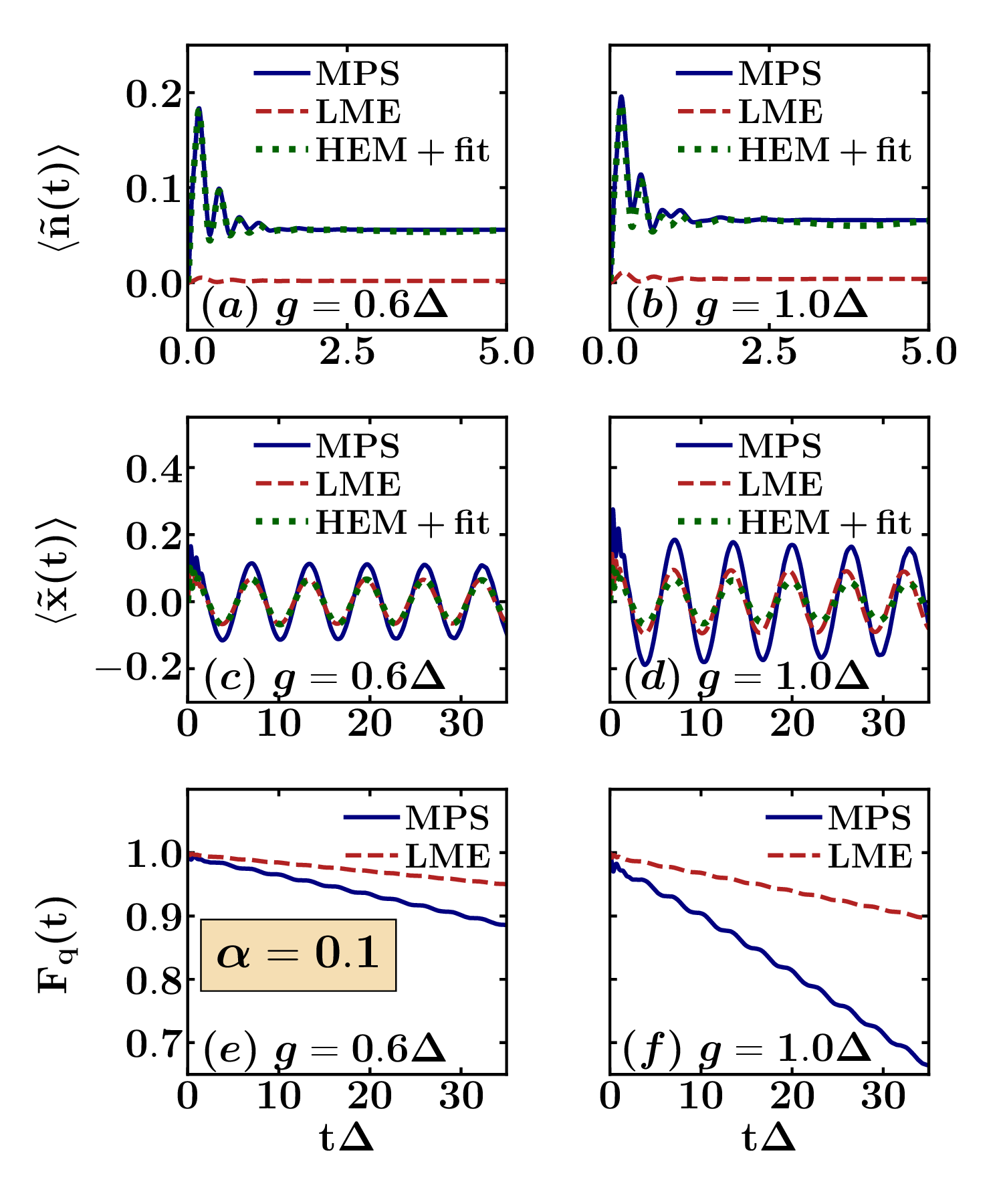}
\caption{\label{fig:xF1_sc} Oscillator number (in panels $(a)$ and $(b)$), oscillator position (in panels $(c)$ and $(d)$) and qubit fidelity (in panels $(e)$ and $(f)$) as a function of time (in units of $1/\Delta$) at $g=0.6\Delta$ and $g=1.0\Delta$, respectively, for $\alpha=0.1$ by using different methods: MPS simulations (blue solid line), the LME solution (red dashed line) and the HEM approach (green dotted line), improved by the fitting of the spin observables as in Eqs.~(\ref{eq:sigmarenx}-\ref{eq:sigmareny}). Parameters: $\Delta=1,\omega_0=10\Delta,g=\{0.6\Delta,1.0\Delta\},\alpha=0.1,\omega_c=50;$ for MPS simulations $N_o=20,N=500,N_{ph}=3,D_{\rm{max}}=20$}
\end{figure}

For $\alpha=0.1$, we find interesting behaviours with increasing $g$. We start again from $g=0.6\Delta$, which has been considered both in Fig.~\ref{fig:FFT_sc} and~\ref{fig:xF01_sc}. We recall that, for $\alpha=0.1$, the $z$-component evaluation is not effective to get the spin component along the $z$-axis. Therefore, we investigate if a Bloch vector evaluation can still be achieved. As shown in Fig.~\ref{fig:xF1_sc}, the time evolution of the oscillator number agrees with that predicted from the HEM improved by the fitting of the spin $z$-component (Eq.~(\ref{eq:sigmarenz})), while the LME remains perturbative and does not significantly change the state from the initial vacuum. For the position, instead, the agreement is good at $g=0.6\Delta$ between LME and HEM, while there is a slightly different amplitude in the MPS position. In fact, LME fails at very large coupling to the bath while HEM at high renormalization of the qubit dynamics. Thus, these perturbative approaches are unable to properly transfer the effect of the environment to the qubit through the interaction with the oscillator. To understand this point, one can consider the mapped model in Eq.~(\ref{eq:mappedH}) where the effective spectral density is Ohmic at low frequencies with coupling proportional to $g^2\alpha/\omega_0^2$. For $g=1.0\Delta$, we observe also a discrepancy between LME and HEM, due to the increased $g$. Moreover, they are phase-shifted with respect to the MPS due to the fact that a large value of the coupling in the mapped model \cite{mapzueco} means a qubit strongly coupled to the effective bath. \\
The panels $(e)$ and $(f)$ of Fig.~\ref{fig:xF1_sc} show qubit fidelity for $\alpha=0.1$, comparing MPS and LME methods. For $g=0.6\Delta$, LME gives an incorrect behaviour, while the MPS simulations provide the limiting case $F_q(5T\Delta)\approx90\%$ in which the Bloch vector evaluation can be performed. Indeed, one can still determine $x$, $y$ and $z$ spin components from the behaviour of oscillator position, momentum and number as predicted by HEM solutions. In panel Fig.~\ref{fig:xF1_sc}$(f)$, for $g=1.0\Delta$, the difference between LME and MPS becomes large, indicative of the fact that in this regime LME fails, while for MPS $F_q(5T\Delta)\approx60\%$. In fact, LME would provide still a perturbative description, with an acceptable fidelity; MPS, however, shows a fidelity almost linearly decreasing so that it results $F_q(5T\Delta)\approx60\%$. Obviously, the qubit fidelity with respect to its free evolution would be strictly $100\%$ only when the HEM
solutions are valid, hence for very low $g$ and $\alpha$. We observe in this intermediate regime between the strong and the ultra-strong coupling that up to our final time of simulations ($t\approx5T=35/\Delta$), it reaches the $90\%$. Moreover, the entanglement in the same intermediate regime becomes very high (see Fig.~\ref{fig:ent}), meaning that the qubit is inextricably linked to the oscillator. For these reasons we choose $90\%$ as the threshold for the fidelity of a reliable qubit state evaluation.\\
Summarizing, we have found that for small $\alpha$ in the strong coupling regime both the $z$-component evaluation for an up state and for a generic state can be achieved. For increasing $\alpha$, anyway, the $z$-component evaluation is unachievable, while for $g\leq0.6$ a full Bloch vector evaluation is still attainable (see Fig.~\ref{fig:diag_fase} and Table~\ref{tab:fid} in the concluding section \ref{sec:conclusioni}).\\
In analogy with weak coupling regime discussed in the previous subsection, we find that the effect of the decoherence helps the full Bloch vector evaluation by reducing the initial transient and the oscillator is in a moving momentum squeezed state as pointed out in \ref{subsec:deco_squeezed}.
%\newpage
\subsection{\label{subsec:USC}Ultra-strong coupling regime - interactions prevent Bloch vector evaluation}
Lastly, with the qubit frequency $\Delta$ taken as unity and $\omega_0=10\Delta$, by focusing on $g/\Delta=5$ and tuning $\alpha\in\{0.01,0.1\}$ such that $g/\Delta\geq0.1\omega_0/\Delta$ (see Fig.~\ref{fig:diag_fase}), the quantum Rabi system is in the ultra-strong coupling regime. \\

\begin{figure}[b]
\centering
\includegraphics[scale=0.4]{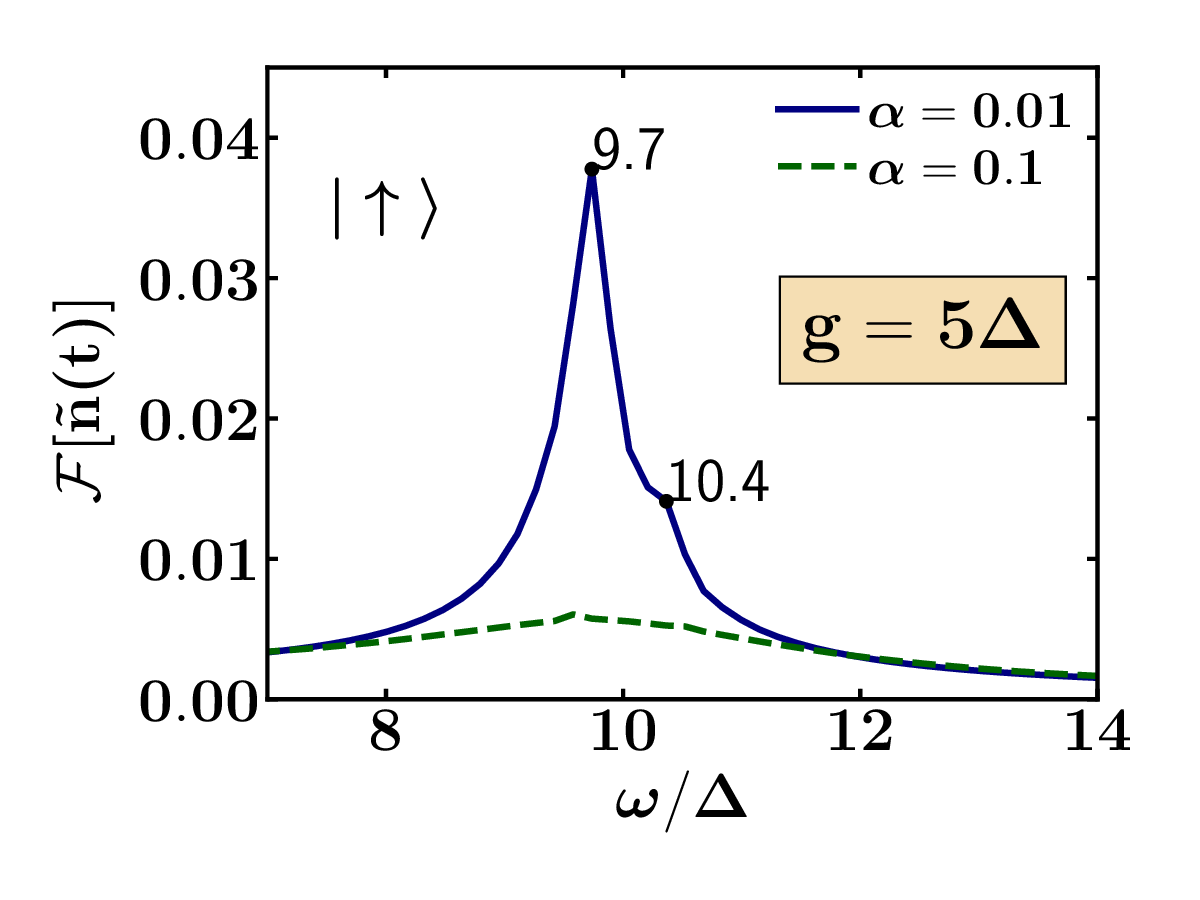}
\caption{\label{fig:FFT_usc} FFT of the mean oscillator number $\tilde{n}$ computed through MPS simulations as a function of the frequency (in units of $\Delta$) for two values of $\alpha$. Parameters: $\Delta=1,\omega_0=10\Delta,g=5\Delta,\alpha=\{0.01,0.1\},\omega_c=50;$ for MPS simulations $N_o=20,N=500,N_{ph}=3,D_{\rm{max}}=20$}
\end{figure}

As in the previous regime, if one chooses $\ket{0}\otimes \ket{\uparrow}$ as pure separable initial state, for low values of $\alpha$ it is possible to follow the peaks of the FFT of the mean oscillator number. In Fig.~\ref{fig:FFT_usc} we see, as expected, that, for $\alpha=0.01$, a peak is prominent, while, for $\alpha\geq 0.1$, the FFT is almost flat. However, even for $\alpha=0.01$, the frequency corresponding to the peak is no more in perfect agreement with the second-order perturbation theory shift given in Eq.~(\ref{eq:shift2ord}) because the system is in the non-perturbative region of ultra-strong coupling regime \cite{rossatto2017spectral} ($g\approx\Delta+\omega_0$). Moreover, the peak corresponding to the down contribution is also emerging. Therefore, the $z$-component evaluation is not feasible because of the marked back-action of the oscillator. Below we will show that, in this coupling regime, qubit and oscillator are indeed strongly entangled.\\

\begin{figure}[t]
\centering
\includegraphics[scale=0.31]{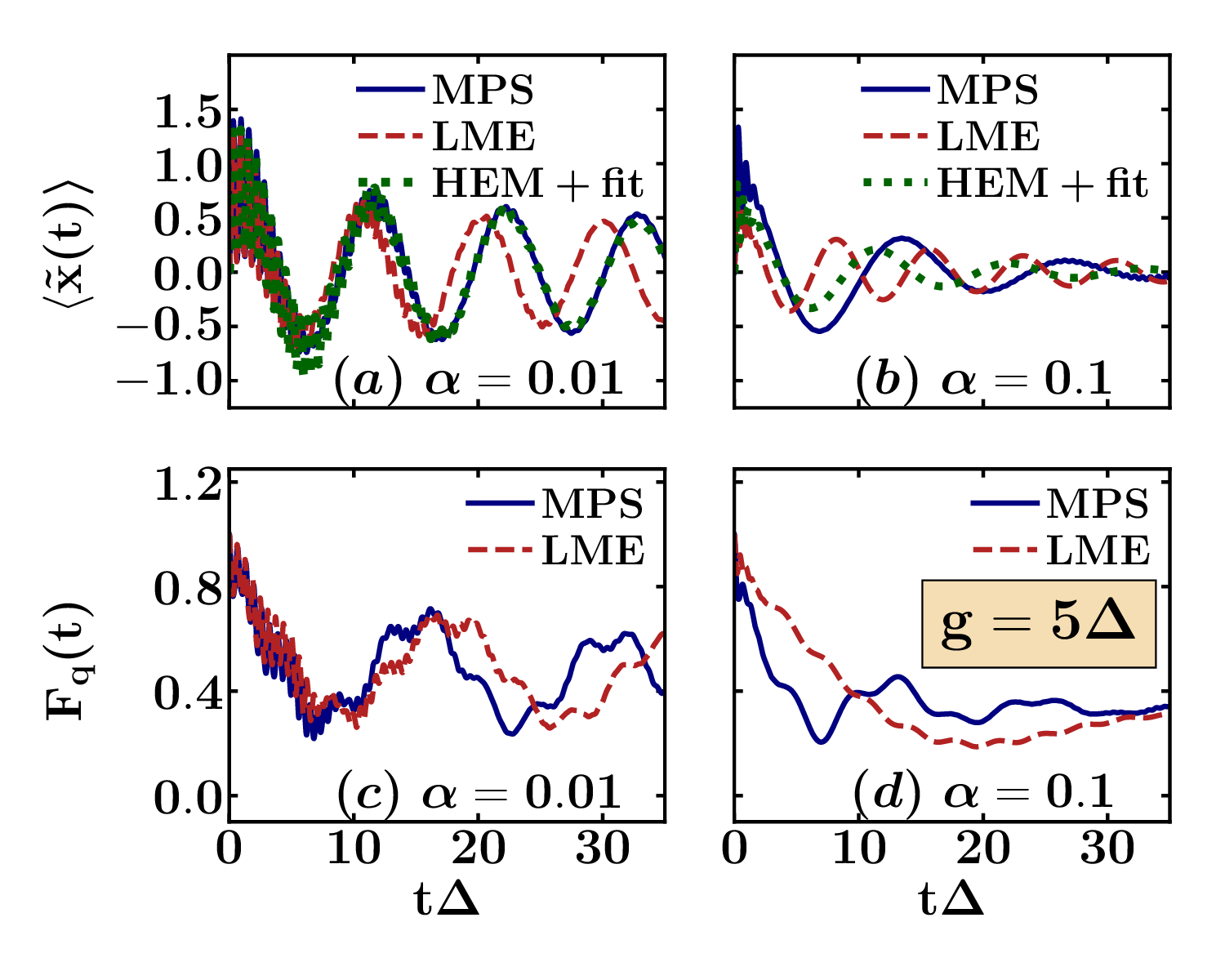}
\caption{\label{fig:xF_usc} Oscillator position (in panels $(a)$ and $(b)$) and qubit fidelity (in panels $(c)$ and $(d)$) as a function of time (in units of $1/\Delta$) for $\alpha=0.01$ and $\alpha=0.1$, respectively, using three methods: MPS simulations (blue solid line), LME solutions (red dashed line) and HEM approach (green dotted line), improved by the fitting of spin observables as in Eqs.~(\ref{eq:sigmarenx}-\ref{eq:sigmareny}). Parameters: $\Delta=1,\omega_0=10\Delta,g=5\Delta,\alpha=\{0.01,0.1\},\omega_c=50;$ for MPS simulations $N_o=20,N=500,N_{ph}=3,D_{\rm{max}}=20$}
\end{figure} 

\begin{figure*}
\centering
\includegraphics[scale=0.55]{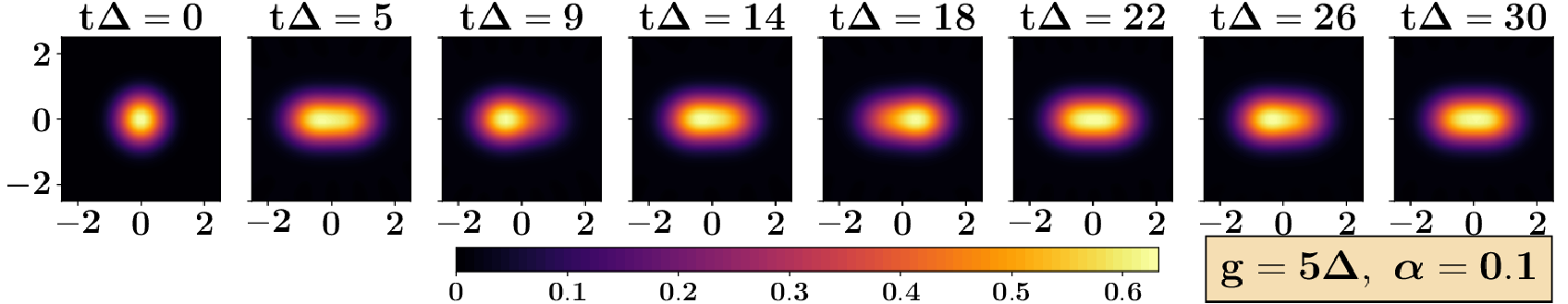}
\caption{\label{fig:wigner_usc} Reduced density matrix of the oscillator computed through MPS simulations at $8$ times $t\Delta$ during its evolution. From this, we obtain the Wigner quasi-probability distribution by using the proper function of the package QuTip \cite{2012qutip}. Parameters: $\Delta=1,\omega_0=10\Delta,g=5\Delta,\alpha=0.1,\omega_c=50;$ for MPS simulations $N_o=20,N=500,N_{ph}=3,D_{\rm{max}}=20$}
\end{figure*}

For the initial state $\ket{0}\otimes\ket{\nearrow}$ and the bath empty, regardless of the value of $\alpha$, Bloch vector evaluation is no more feasible (see Fig.~\ref{fig:xF_usc}). In particular, the agreement between different computational methods is good for $\alpha=0.01$ at short times, while there is a dephasing in LME solution due to the fact that a large value of the coupling in the mapped model \cite{mapzueco,de2022signatures} means a qubit strongly coupled to the effective bath. For larger $\alpha$ also HEM fails because the fit is not enough to describe the qubit dynamics. In Fig.~\ref{fig:xF_usc}$(c)$ and Fig.~\ref{fig:xF_usc}$(d)$ we compare the fidelity for the MPS and LME approaches. For $\alpha=0.01$, the results are very similar, oscillating around $0.5$ due to the growing entanglement, while for $\alpha=0.1$ they have an exponential decaying behaviour, with an overmodulation for MPS. Anyway, during the first $5$ qubit periods, both MPS and LME solutions give a minimum fidelity close to $20\%$. We emphasize that a Bloch vector evaluation is still possible, at least for small $\alpha$ values, but what we are following is the qubit state as a part of the coupled system. As a consequence, it is very entangled, so its evolution is far from the free one and the fidelity is small (see Fig.~\ref{fig:diag_fase} and Table~\ref{tab:fid} in the concluding section \ref{sec:conclusioni}). \\

This set of parameters allows us to observe the moving of the squeezed state (being the amplitude of the displacement proportional to $\bar{g}$) (Fig.~\ref{fig:wigner_usc}). The position of the oscillator oscillates around zero as shown in Fig.~\ref{fig:xF_usc}, while its momentum remains close to the zero. The oscillator starts in the vacuum and evolves in a momentum squeezed state whose center oscillates with damping around zero until stopping at the origin of the quantum phase space, as underlined in \ref{subsec:deco_squeezed}. 
%\newpage

\section{Discussions and conclusions}\label{sec:conclusioni}
In the present work, we have provided an exhaustive theoretical analysis of a method for evaluating the coordinates of the qubit Bloch vector through the dynamics of its coupled oscillator. In particular, we have tested the functional relationships derived from the perturbative (in $g$) solutions of HEM, by simulating the quantum Rabi model with MPS numerical method. We have compared it also with a numerical solution of the LME, which is perturbative in the oscillator-bath coupling and unable to describe the non-Markovian effects of the bath. We have found the critical values of the coupling to the bath for the $z$-component evaluation in the cases of a pure initial up state and a generic qubit state on the Bloch surface as consequences of the coupling to the bath. Moreover, we have implemented full Bloch vector evaluation, by computing the time evolution of the oscillator observables. We have found the parameter regimes where the procedure does not work, taking the qubit fidelity with respect to its free evolution as a measure of the quality of the Bloch vector evaluation. Table~\ref{tab:fid} shows the qubit fidelity computed at time $t_{\rm{final}}=35/\Delta$ for weak and strong coupling regimes, while for the ultra-strong regime the minimum value is reached at time $t=7/\Delta$ for the MPS method and $t=19.5/\Delta$ for the LME one.

\begin{table}[ht]%The best place to locate the table environment is directly after its first reference in text
\caption{\label{tab:fid}Minimum qubit fidelity $F_q(t)$ (\%) for the three regimes analysed in the paper: weak coupling (WC), strong coupling (SC) and ultra-strong coupling (USC)
}
\begin{center}
\begin{tabular}{l|ccc}
\hline
\hline
\textrm{Method}&
\textrm{WC}&
\textrm{SC}&
\textrm{USC}\\
\textrm{}&
\textrm{$g=0.1\Delta$}&
\textrm{$g=0.6\Delta$}&
\textrm{$g=5\Delta$}\\
\textrm{}&
\textrm{$\alpha=0.05$}&
\textrm{$\alpha=0.1$}&
\textrm{$\alpha=0.1$}\\
\hline
MPS & 99.8 & 61.2 & 20.4 \\
LME & 99.9 & 88.0 & 18.7 \\
\hline
\hline
\end{tabular}
\end{center}
\end{table}

\begin{figure}[t]
\centering
\includegraphics[scale=0.25]{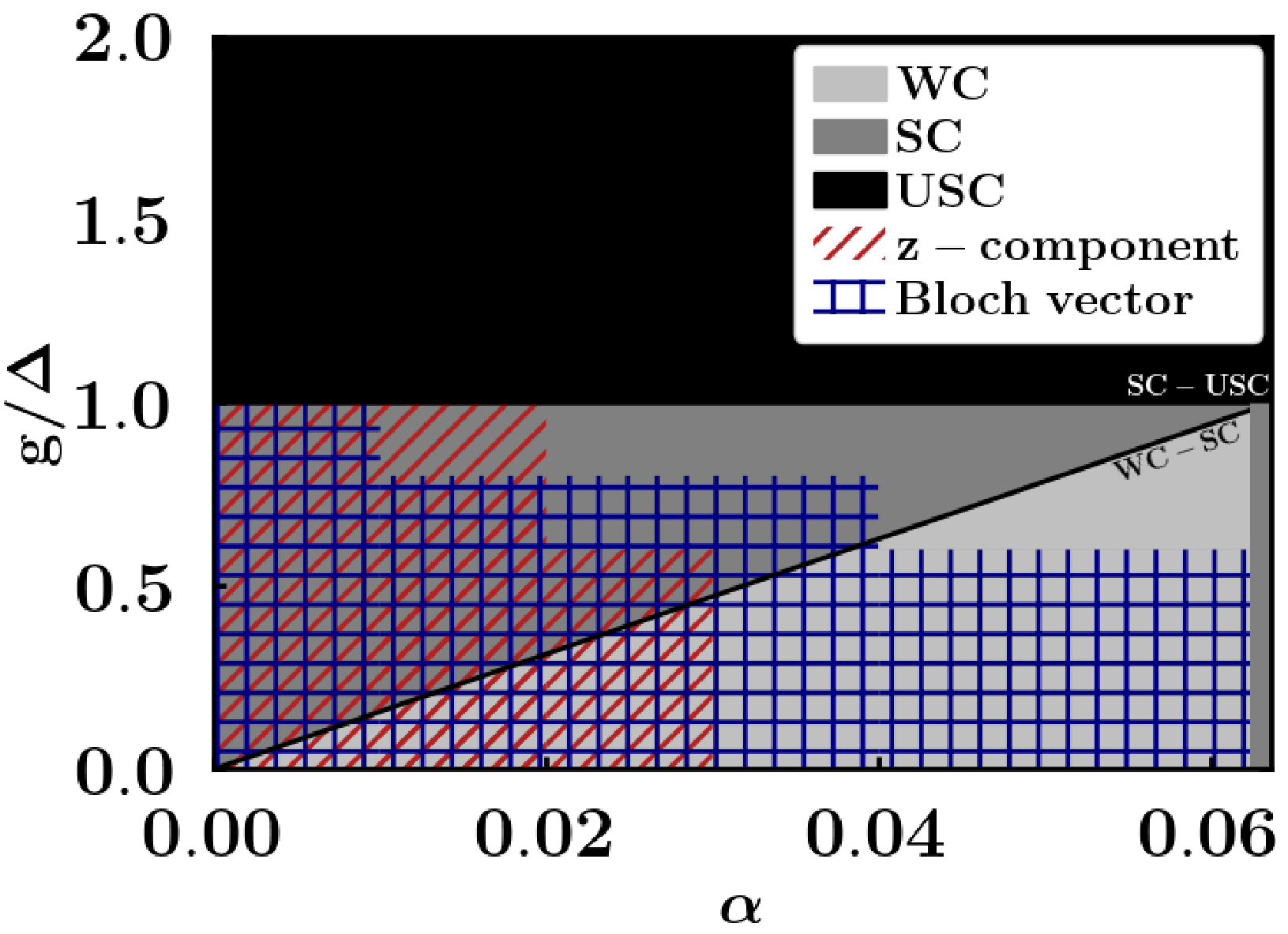}
\caption{\label{fig:diag_fase}Phase diagram of the quantum Rabi model with $g$ vs. $\alpha$. On the background there are the three regions of parameters analysed in the paper: weak coupling (WC, light grey), strong coupling (SC, grey) and ultra-strong coupling (USC, black). The two methods under study are shown using different patterns: $z$-component evaluation with red ``//" and Bloch vector one with blue ``+". The grey vertical line at $\alpha=2/(10\pi)$ underlines that above this value the regime is again the strong coupling one because $\Gamma/\Delta>0.1\omega_0/\Delta$. Parameters: $\Delta=1,\omega_0=10\Delta,g\in[0\Delta,2\Delta],\alpha\in[0,2/(10\pi)],\omega_c=50, N_o=20,N=500,N_{ph}=3,D_{\rm{max}}=20$}
\end{figure}

In Fig.~\ref{fig:diag_fase} we show the validity of the two methods used for the $z$-component evaluation and the full Bloch vector evaluation via oscillator dynamics in the three regimes: weak, strong and ultra-strong coupling. We stop the plot at $\alpha=2/(10\pi)$ that corresponds to $\Gamma/\Delta=0.1\omega_0/\Delta$. The latter indicates that the estimation of the perturbative decay rate $\Gamma$ for the definition of the parameter regions no longer holds. Hence, above this value of $\alpha$, the regime is again the strong coupling one. We emphasize that for internal coupling $g\leq0.8\Delta$ the full state evaluation is possible for coupling strengths $\alpha$ bigger than those necessary for the $z$-component evaluation. Nevertheless, in the weak coupling regime the procedure with number of quanta is still the most sensitive to the initial state of the qubit. Furthermore, for increasing internal couplings, even if the LME solution would give a reliable qubit fidelity, regardless of the coupling to the bath, the MPS simulations have shown that it is already much smaller than the acceptance threshold of $90\%$. Nevertheless, we are still following the qubit, but the interactions have modified significantly its dynamics with respect to the free one and the method fails. In the ultra-strong regime the fidelity oscillates around the value of $50\%$, that can be easily interpreted by looking at the qubit entanglement in time. 

\begin{figure}[ht]
\centering
\includegraphics[scale=0.26]{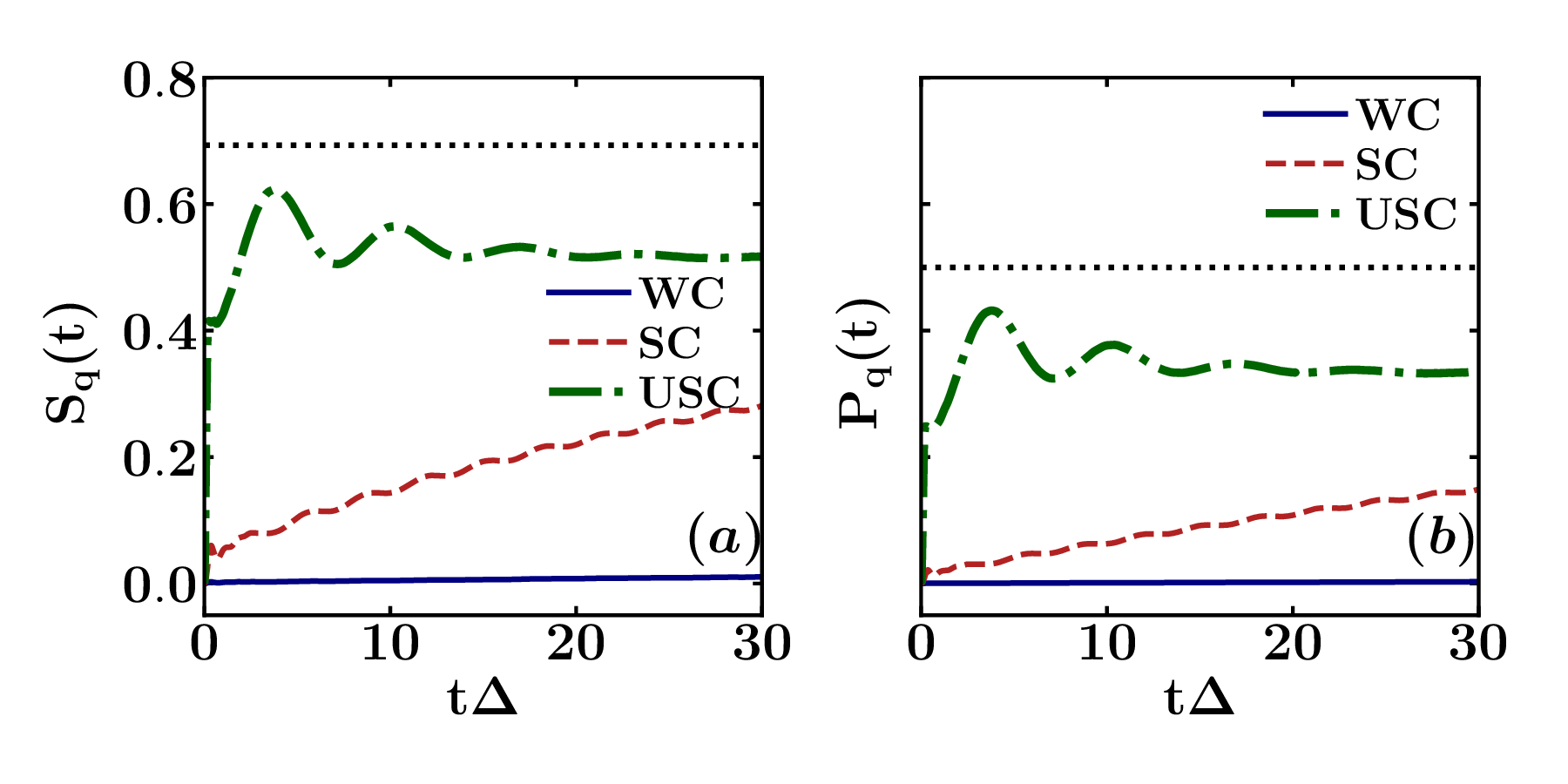}
\caption{\label{fig:ent}Qubit Von Neumann entropy ($S_q=-\Tr\{\rho_q \log\rho_q\}$) and purity ($P_q=1-\Tr\{\rho_q^2\}$) computed through MPS simulations in panels $(a)$ and $(b)$ respectively, for $g=0.1\Delta,\,\alpha=0.05$, i.e. weak coupling (blue solid line), $g=0.6\Delta,\,\alpha=0.1$ i.e. strong coupling (red dashed line) and $g=5\Delta,\,\alpha=0.1$ i.e. ultra-strong coupling (green dash-dotted line). Parameters: $\Delta=1,\omega_0=10\Delta,g=\{0.1\Delta,0.6\Delta,5\Delta\},\alpha=\{0.05,0.1\},\omega_c=50;$ for MPS simulations $N_o=20,N=500,N_{ph}=3,D_{\rm{max}}=20$}
\end{figure}

In Fig.~\ref{fig:ent}, the qubit Von Neumann entropy ($S_q=-\Tr\{\rho_q \log\rho_q\}$) and the purity ($P_q=1-\Tr\{\rho_q^2\}$) are displayed for the three different regimes of parameters examined in the paper. We clearly observe how they increase towards the maximum values (dotted black lines $S_q^{max}=\log 2$ and $P_q^{max}=1/2$, respectively) in the ultra-strong coupling regime. Actually, for the coupled system qubit-oscillator, entanglement becomes bigger and bigger. This is the main reason why the qubit fidelity with respect to its free evolution becomes smaller and smaller.\\

\begin{figure}[b]
\centering
\includegraphics[scale=0.35]{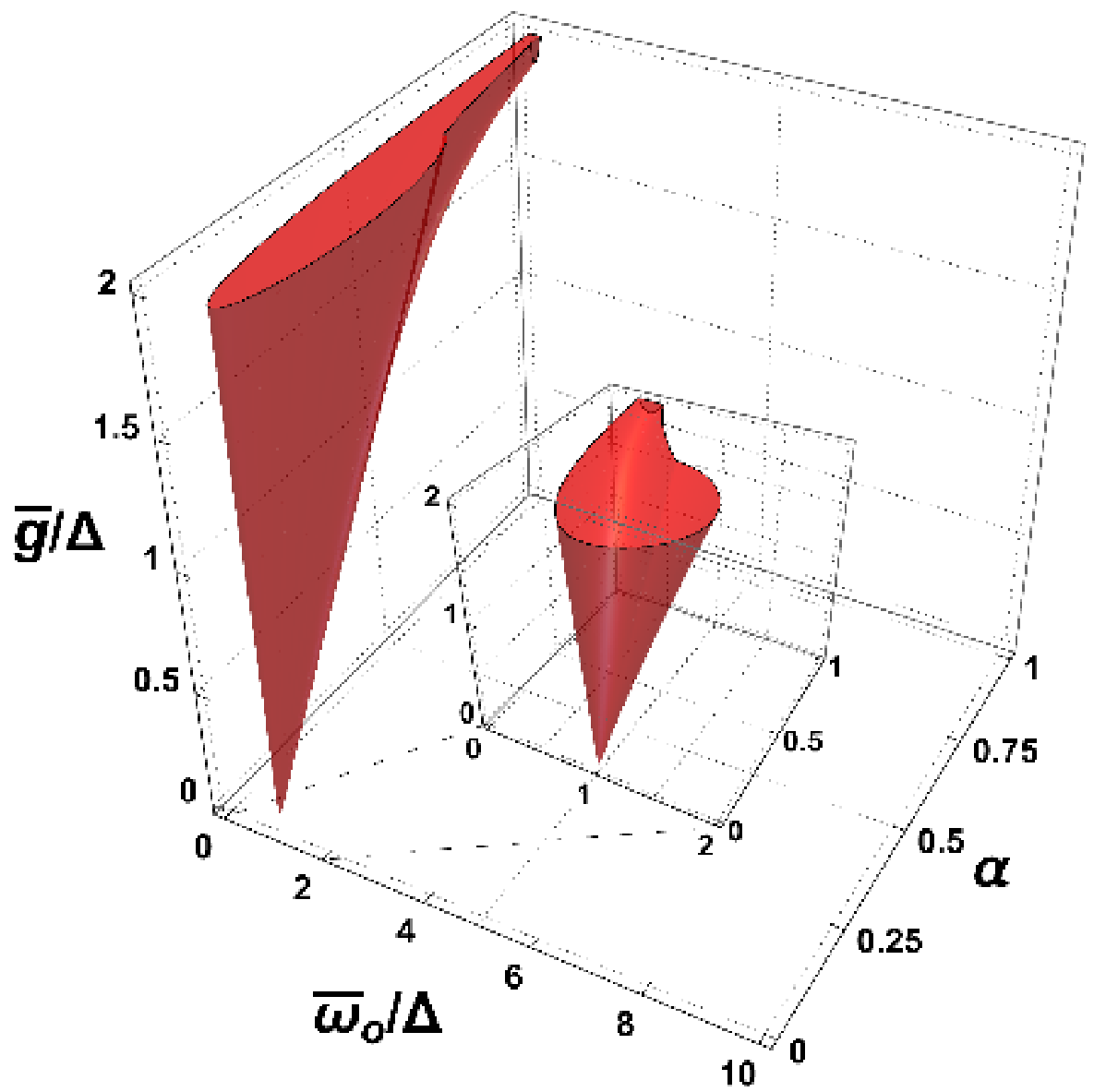}
\caption{\label{fig:diagfase1}Bloch vector evaluation is not valid for the values of ($\bar{\omega}_0/\Delta,\alpha,\bar{g}/\Delta$) in the red 3D region, by taking $\Delta$ as unity. The region is the result of the union among the inequalities (Eqs.~\ref{eq:diseg_x1}-\ref{eq:diseg_p2}) that define the conditions for obtaining (Eqs.~\ref{eq:partsolx}-\ref{eq:partsolp}) for the oscillator dynamical variables}
\end{figure}

Finally, we point out how the procedure analysed in our work to evaluate the qubit state is not restricted to the anti-adiabatic regime ($\Delta\ll\bar{\omega}_0$), but the description is valid for a wider range of parameters. In Fig.~\ref{fig:diagfase1}, we depict the forbidden region of the set of three parameters ($\bar{\omega}_0/\Delta,\alpha,\bar{g}/\Delta$), while $\Delta=1$, by limiting the $\bar{g}/\Delta$ values to $2$ and the $\alpha$ ones to $1$ (very large coupling strengths) and $\bar{\omega}_0/\Delta$ to $10$, of the order of the value chosen in the paper. This region is obtained by performing the union among the inequalities that define the conditions for neglecting the spurious terms and obtaining the analytical particular solutions for $\tilde{x}(t)$ and $\tilde{p}(t)$ (Eqs.~\ref{eq:partsolx}-\ref{eq:partsolp}). For the mean position, at fixed initial conditions $\langle\sigma_x(0)\rangle$ and $\langle\sigma_y(0)\rangle$ the inequalities to be both verified read:
\begin{align}\label{eq:diseg_x1}
    \frac{2\pi\alpha\Delta\bar{\omega}_0^2\bar{g}\langle\sigma_x(0)\rangle}{\left(\bar{\omega}_0^2-\Delta^2\right)^2+\pi^2\alpha^2\bar{\omega}_0^2\Delta^2}&\ll 1\\\label{eq:diseg_x2}
    \frac{2\pi\alpha\Delta\bar{\omega}_0^2\bar{g}\langle\sigma_y(0)\rangle}{\left(\bar{\omega}_0^2-\Delta^2\right)^2+\pi^2\alpha^2\bar{\omega}_0^2\Delta^2}&\ll 1,
\end{align}
and similarly those for the momentum are:
\begin{align}\label{eq:diseg_p1}
    \frac{2\pi\alpha\Delta^2\bar{\omega}_0\bar{g}\langle\sigma_x(0)\rangle}{\left(\bar{\omega}_0^2-\Delta^2\right)^2+\pi^2\alpha^2\bar{\omega}_0^2\Delta^2}&\ll 1\\\label{eq:diseg_p2}
    \frac{2\pi\alpha\Delta^2\bar{\omega}_0\bar{g}\langle\sigma_y(0)\rangle}{\left(\bar{\omega}_0^2-\Delta^2\right)^2+\pi^2\alpha^2\bar{\omega}_0^2\Delta^2}&\ll 1.
\end{align}

Moreover, our analysis is valid for a wider range of temperatures. In fact, since that we do not describe the measurement apparatus, we analyse the effect of thermal photon population on the number of quanta in the oscillator by introducing a non-zero temperature bath. We observe that up to temperatures corresponding to $\Delta$ in the weak coupling regime, where the functional relationships are strictly valid, our results for $T=0$ are robust (for more details, see App.~\ref{app:temp}).

At this stage, as recalled in the introduction \ref{sec:intro}, we can thoroughly discuss the results of the experiments in Ref.~\onlinecite{ficheux_tomo}. Indeed, in these experiments, a full qubit state reconstruction is realised by reading the quadratures of a coupled oscillator via homodyne and heterodyne measurements. The unraveling of quantum trajectories has been used in order to follow the qubit dynamics. The method is also known as the method of quantum jumps, exactly the same as the Lindblad operators \cite{breuer}. The resulting master equation is then an appropriately weighted stochastic average over all of the different times at which the jumps could occur, and all of the different types of jumps that can occur \cite{daley2014quantum}. Hence, this approach is equivalent to the Lindblad approach for the qubit in the limit of a sufficiently large number of experimental realizations. It is remarkable that a full quantum state reconstruction can be obtained at any time by raw averaging measurement outcomes of many realizations of a single experiment despite the incompatibility of the three spin components of the qubit.\\
In our work the idea of how performing the Bloch vector evaluation is similar to the qubit state reconstruction in Ref.~\onlinecite{ficheux_tomo}, that is to read the oscillator dynamics. In fact, the functional relationships linking the qubit dynamics to the oscillator evolution, that we have derived from HEM, bear some resemblance to those used in Ref.~\onlinecite{ficheux_tomo} to interpret measurements.We explain the conceptual differences with respect to our approach in more detail below. In this work, at odds with Ref.~\onlinecite{ficheux_tomo}, we have exploited the effects of the bath to anticipate the Bloch vector evaluation. Moreover, the LME solved in this work takes qubit and oscillator on the same footing. Finally, we have introduced dissipation and decoherence through the presence of an Ohmic bath, without describing a measurement apparatus. Our model is clearly different from that used in the experiment. Indeed, we do not use pulses to drive dispersively the system and measure the quadratures. We are able to observe the shift of the oscillator frequency depending on the initial qubit state only from the number of quanta in the harmonic oscillator. This is possible because the oscillator number has an initial transient oscillating with the shifted frequencies. In particular, in Eq.~\ref{eq:NHEM}, these oscillations are due to the variances of $\tilde{x}$ and $\tilde{p}$ that include terms at the second-order in the coupling $g$. Moreover, transient at smaller times means higher frequencies, therefore,
one can get some information about the degree of freedom with the highest energy: the oscillator. In any case, if we increase $g$, we can observe the shifts of the oscillator frequency also in the oscillator quadratures. This behaviour is analogous to that experimentally measured by means of quadratures. In fact, the effect of the oscillator drive in experiments can be roughly interpreted as an increase, even temporarily, of the oscillator coupling with other degrees of freedom. A generalization of our study to a more realistic model, including the measurement process description, deserves future investigations \cite{breuer,walls1985analysis}. \\ 

The study carried out in the present work can be extended to the systems where there are more qubits and more oscillators. However, while these generalizations are straightforward in HME and LME approaches, they would be clearly more challenging within a MPS framework. Such architectures are typical of the recent noisy intermediate-scale quantum processors. Furthermore, the oscillator could be investigated in more detail, since the interest about the bosonic behaviour, specially in circuit QED, is growing. In fact, quantum information can be encoded into subspaces of a bosonic superconducting cavity mode with long coherence time \cite{ma2021quantum}. 

\bmhead{Acknowledgments}
We acknowledge interesting discussions with R. Fazio. A.N. acknowledges support from the Max Planck-UBC-UTokyo Center for Quantum Materials and the Canada First Research Excellence Fund (CFREF) Quantum Materials and Future Technologies Program of the Stewart Blusson Quantum Matter Institute (SBQMI).
% REV\TeX{}, offering suggestions and encouragement, testing new versions,
% \dots
\bmhead{Data Availability Statement}
The datasets generated and analyzed in this study are available upon reasonable request from the corresponding author. This
manuscript has associated data in a data repository.[Authors’ comment: All data used in this publication are available on reasonable request by contacting
the corresponding author.]
\bmhead{Declarations}
\bmhead{Conflict of interest}
The authors declare that they have no competing interests.

% \bigskip

% Springer journals and proceedings: \url{https://www.springer.com/gp/editorial-policies}

% Nature Portfolio journals: \url{https://www.nature.com/nature-research/editorial-policies}

% \textit{Scientific Reports}: \url{https://www.nature.com/srep/journal-policies/editorial-policies}

% BMC journals: \url{https://www.biomedcentral.com/getpublished/editorial-policies}
% \end{flushleft}

\begin{appendices}

\appsection{Perturbative solution of closed Rabi model}\label{app:closedRabi}
We solve the closed Rabi model (\ref{eq:closedH}), for low values of the coupling $g$ by restricting to the $4$-dimensional subspace of the Hilbert space spanned by $\{\ket{\downarrow,0},\ket{\uparrow,0},\ket{\downarrow,1},\ket{\uparrow,1}\}$. The resulting perturbative Rabi Hamiltonian is block-type and we can diagonalise it block by block, by obtaining two pairs of eigenstates and eigenvalues. For the block identified by $\{\ket{\downarrow,0},\ket{\uparrow,1}\}$ the eigenvalues are $\epsilon_{\pm}=\frac{\omega_0}{2}\pm \Omega_{CR}$, where $\Omega_{CR}=\sqrt{g^2+((\omega_0+\Delta)/2)^2}$ is the oscillation frequency due to the counter rotating terms of the Rabi interaction and the respective eigenvectors read 
\noindent
\begin{equation}
\begin{cases}
    \ket{e_{CR}}=\sin{\frac{\theta_{CR}}{2}}\ket{\downarrow,0}+\cos{\frac{\theta_{CR}}{2}}\ket{\uparrow,1}\\
    \ket{g_{CR}}=-\cos{\frac{\theta_{CR}}{2}}\ket{\downarrow,0}+\sin{\frac{\theta_{CR}}{2}}\ket{\uparrow,1},
\end{cases}\\
\end{equation}
with $\ket{e_{CR}}$ ($\ket{g_{CR}}$) which has the higher (lower) energy $\epsilon_+$ ($\epsilon_-$) and $\theta_{CR}=\arctan{\left(\frac{2g}{\Delta+\omega_0}\right)}$. Similarly for the block identified by $\{\ket{\uparrow,0},\ket{\downarrow,1}\}$ where the eigenvalues are given by $\xi_{\pm}=\frac{\omega_0}{2}\pm \Omega_{R}$, where $\Omega_{R}=\sqrt{g^2+((\omega_0-\Delta)/2)^2}$ is the oscillation frequency due to the rotating terms of the Rabi interaction, those of the Jaynes-Cummings model, and the eigenvectors are
\noindent
\begin{equation}
\begin{cases}
    \ket{e_{R}}=\sin{\frac{\theta_{R}}{2}}\ket{\downarrow,0}+\cos{\frac{\theta_{R}}{2}}\ket{\uparrow,1}\\
    \ket{g_{R}}=-\cos{\frac{\theta_{R}}{2}}\ket{\downarrow,0}+\sin{\frac{\theta_{R}}{2}}\ket{\uparrow,1},
\end{cases} 
\end{equation}
with $\theta_{R}=\arctan{\left(\frac{2g}{\omega_0-\Delta}\right)}$. Starting from the initial state $\ket{\psi_0}=\alpha\ket{\uparrow,0}+\beta\ket{\downarrow,0}=\cos{\theta/2}\ket{\uparrow,0}+\sin{\theta/2}e^{-i\phi}\ket{\downarrow,0}$ we can compute the time evolution of the observables of both the qubit and the oscillator and in particular we find for the oscillator number:
\begin{align}
    \langle\tilde{n}(t)\rangle&=\left\|\beta\right\|^2\sin^2{\theta_{CR}}\sin^2{(\Omega_{CR}t)}\\\nonumber
    &+\left\|\alpha\right\|^2\sin^2{\theta_{R}}\sin^2{(\Omega_{R}t)},
\end{align}
which by performing the Fourier transform gives 
\begin{eqnarray}\nonumber
    \mathcal{F}[\langle\tilde{n}(t)\rangle](\nu)=&&\left\|\beta\right\|^2\frac{\sin^2{\theta_{CR}}}{4}[2\delta(\nu)+\pi(\delta(\Omega_{CR}\\\nonumber
    &&+\pi\nu)-\delta(\Omega_{CR}-\pi\nu))]\\\nonumber
    &&+\left\|\alpha\right\|^2\frac{\sin^2{\theta_{R}}}{4}[2\delta(\nu)+\pi(\delta(\Omega_{R}\\
    &&+\pi\nu)-\delta(\Omega_{R}-\pi\nu))].
\end{eqnarray}
Analogously, we find for the magnetization
\begin{eqnarray}
    \langle\sigma_{z}(t)\rangle=&&\left\|\alpha\right\|^2(\sin^2{\theta_{R}}\cos{(2\Omega_{R}t)}+\cos^2{\theta_{R}})\\\nonumber
    &&-\left\|\beta\right\|^2(\sin^2{\theta_{CR}}\cos{(2\Omega_{CR}t)}+\cos^2{\theta_{CR}}),
\end{eqnarray}
and its Fourier transform reads
\begin{eqnarray}\nonumber
    \mathcal{F}[\langle\sigma_{z}(t)\rangle](\nu)=&&-\left\|\beta\right\|^2[\cos^2{\theta_{CR}}\delta(\nu)+(\delta(\Omega_{CR}\\\nonumber
    &&+\pi\nu)+\delta(\Omega_{CR}-\pi\nu))\frac{\pi}{2}\sin^2{\theta_{CR}}]\\\nonumber
    &&+\left\|\alpha\right\|^2[\cos^2{\theta_{R}}\delta(\nu)+(\delta(\Omega_{R}+\pi\nu)\\
    &&+\delta(\Omega_{R}-\pi\nu))\frac{\pi}{2}\sin^2{\theta_{R}}].
\end{eqnarray}
We observe that the mean oscillator number in time oscillates with the same frequencies of the magnetization and in fact the Fourier transform shows peaks at the frequencies $\omega_0\pm\Delta$ for $g\rightarrow 0$. We emphasize that without taking into account the counter rotating terms, we cannot predict an oscillation at frequency $\Omega_{CR}$ and hence a peak in the Fourier transform related to the contribution of the down qubit state.\\
The $x$ and $y$ components of the Bloch vector in time are closer to those in the free case (Rabi oscillations at frequency $\Delta$), justifying the choice of solving the coupled equations of motion by substituting the time evolution of Pauli matrices at $g=0$ in HEM approach. For the position and momentum of the oscillator we observe the occurrence of oscillations at frequencies $\omega_0$ and $\Delta$, as an indication of the possibility of following the qubit through the oscillator dynamics. By computing the Fourier transform of $x$ and $y$ components and position and momentum we find also oscillations at frequencies $\omega_0 \pm \Omega_R$ and $\omega_0 \pm \Omega_{CR}$. The amplitudes of these peaks are negligible in the closed system and more and more flattened in the open one. For this reason we cannot observe the shift of the oscillator frequency depending on the qubit state from the quadratures, as usually done in the experiments. Instead, the oscillator number shows the splitting of the frequency due to due variances of position and momentum that involve terms at the higher orders in the coupling $g$.

\appsection{Lindblad master equation}\label{app:Lind}
The most straightforward approach to the dynamics of an open quantum system is based on the solution of a quantum master equation which generalises the Liouville - Von Neumann equation for the reduced density operator of the system of interest. In fact, when the system is \textit{isolated}, described by a time-independent Hamiltonian, or \textit{closed}, described by a time-dependent Hamiltonian (e.g. a driving field), its dynamics is fully characterised by the \textit{Liouville - Von Neumann equation}:
\begin{equation}
	\frac{d\rho_I(t)}{dt}=-\frac{i}{\hbar}[\mathcal{H}_I(t),\rho_I(t)]=\mathcal{L}\rho_I(t),
\end{equation}
written in the interaction picture. The subscript $I$ stays for ``interaction'' and the whole system Hamiltonian is $\mathcal{H}(t)=\mathcal{H}_0+\mathcal{H}_I(t)$, with $\mathcal{H}_0$ the unperturbed energy of the separated quantum systems and $\mathcal{H}_I(t)$ the interaction between them. Moreover, $\mathcal{L}$ is the Liouville superoperator that acting on the density matrix gives its time evolution.\\ 
The famous Lindblad master equation \cite{lindblad} (LME) is a Markovian quantum master equation that has the advantage of being easy to solve numerically through suitable tools, and analytically in some special situations. An open quantum system is said to be Markovian when its behaviour at a given time is independent of its behaviour in the past. The main drawback of Markovian framework is the lack of the back-action upon the system of interest due to the environment, that is indeed neglected due to the hypothesis of the so-called Born-Markov approximation, valid for systems weakly coupled to the environment \cite{derivlindblad}. In fact, according to this approximation, the environment relaxes before the system changes its state, $\rho(t)=\rho_S(t)\otimes\rho_B(0)$ so that it cannot act on the system (\textit{Born approximation}). Furthermore, \textit{Markov approximation}, i.e. $\tau_R\gg\tau_B$, where $\tau_R$ is the relaxation time of the system $S$ and $\tau_B$ is the correlations time of the bath, means that the dynamics over a time of the order of $\tau_B$ cannot be resolved. The Born-Markov approximation does not warrant the complete positivity of the dynamical map. Therefore, we need to perform one further approximation, the \textit{secular or rotating wave approximation (RWA)}. If $\tau_S$ is the time scale of the system evolution and $\tau_S\ll\tau_R$, the non-secular terms may be neglected, since they oscillate very rapidly during the time over which $\rho_S$ varies appreciably. If we separate the Hermitian and non-Hermitian parts of the dynamics of the system, and we return to the Schr\"odinger picture diagonalising the matrix formed by the rate coefficients, we obtain the LME for $\rho_S(t)$:
\begin{equation}
\begin{split}
	&\frac{d\rho_S}{dt}\equiv\mathcal{L}\rho_S=-\frac{i}{\hbar}[\mathcal{H}_S+\mathcal{H}_{LS},\rho_S]\\
	&+\sum_{\omega,k} \gamma_{k}(\omega)\left[L_k(\omega)\rho_S L_k^{\dagger}(\omega)-\frac{1}{2}\left\{L_k^{\dagger}(\omega)L_k(\omega),\rho_S\right\}\right],
	\label {eq:lindblad}
\end{split}
\end{equation}
where the first term is the unitary evolution with $\mathcal{H}_{LS}$ the \textit{Lamb and Stark shift Hamiltonian}, whose role is to renormalize the system energy levels due to the interaction with the environment. The sum is the \textit{Dissipator} with rates $\gamma_k$ and $L_k$ the Lindblad or jump operators. 

\begin{figure*}[t]
\centering
\includegraphics[scale=0.55]{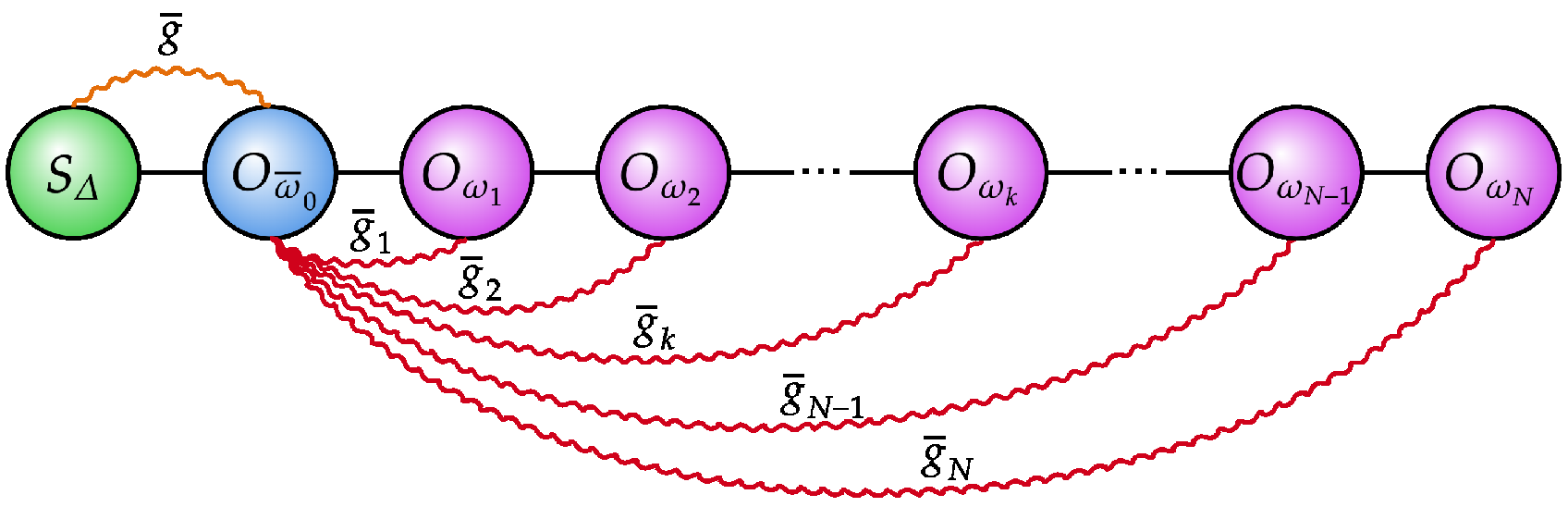}
\caption{\label{fig:MPS}MPS chain of sites representing the Hamiltonian in Eq.~(\ref{eq:Hamiltonian}). The first site is occupied by the qubit, the second one by the oscillator with Hilbert space of dimension $N_o$ and the sites from the third to the $(N+2)$\_th form the Ohmic bath of bosonic modes, each with Hilbert space of dimension $N_{ph}$. Qubit and oscillator are coupled via $\bar{g}$, while the oscillator and the N bath modes are coupled by long range interactions of strength $\bar{g}_j$}
\end{figure*}

When dealing with multipartite systems as in our Hamiltonian given in Eq.~(\ref{eq:Hamiltonian}), the key point is how to eigendecompose the system observables occurring in the interaction Hamiltonian. One way (Local) consists in neglecting the ``internal" interaction among the parties \cite{localglobal}. In this case the eigenstates are factorised and one can only add the dissipators pertaining the different operators in order to describe the dissipation. The other one (Global) instead is based on the use of the correct eigenstates of the Hamiltonian of the system surrounded by the bath. %(Fig.~(\ref{fig:globloc})). 
Moreover, it is known that the global approach provides an accurate description of the system evolution at long timescales \cite{farina2020going}. Thus, in the paper we follow this approach, because we aim to properly describe the interaction between qubit and oscillator up to times of the order of some qubit periods. Hence, we diagonalise the Hamiltonian of qubit plus oscillator system (Eq.~(\ref{eq:Hamiltonian}) without the bath) at each time step $dt$, we eigendecompose the system observables and we evaluate the Lindbladian $\mathcal{L}$. Eq.~(\ref{eq:lindblad}) for the reduced density matrix of the bipartite system $\rho_S(t)$, vector in the Liouville space, is solved numerically through the very well-known fourth-order Runge-Kutta algorithm, derived from two different Taylor expansions of the dynamical variable $\rho_S$ and its derivative \cite{taopang}:
\begin{subequations}
\begin{equation}
    \rho_S(t + dt ) = \rho_S(t) +\frac{1}{6}(c_1 + 2c_2 + 2c_3 + c_4),
\end{equation}
\vspace{0.01cm}
\begin{align}
    c_1 = \,\,&dt\, \mathcal{L}\left(\rho_S(t)\right),\nonumber\\
    c_2 =\,\, &dt\, \mathcal{L}\left(\rho_S(t) + \frac{c_1}{2}\right),\nonumber\\
    c_3 =\,\,&dt\, \mathcal{L}\left(\rho_S(t) + \frac{c_2}{2}\right),\nonumber\\
    c_4 = \,\,&dt\,
    \mathcal{L}\left(\rho_S(t) + c_3\right). 
\end{align}
\end{subequations}

We study the solution of LME in order to compare a Markovian description with other analytical and numerical methods, such as the MPS approach introduced in App.~\ref{app:mps}.

\appsection{Matrix Product States simulations}\label{app:mps}

We solve the Hamiltonian in Eq.~(\ref{eq:Hamiltonian}) using time-dependent MPS simulations. In particular, we adopt the star geometry depicted in Fig.~\ref{fig:MPS} to describe the long-range interactions between the qubit, the harmonic oscillator, and the bath modes. \\
Because of the long-range character of the interactions, we adopt two different methods for the solution of the time-dependent Schr\"odinger equation.\\ 
The first one was developed in Ref.~\onlinecite{longrange} and consists in a first order approximation of the
unitary time-evolution operator in terms of a Matrix Product Operator (MPO). This method, to which we refer to $W^I$ in the following, has an error per site 
which diverges with the system size $L$, while giving a time-step error of $O(dt^2)$.\\
The second method we use is the time-dependent variational principle (TDVP) \cite{Haegeman2011,Haegeman2016,Paeckel2019}, where the time-dependent Schr\"odinger equation is projected to the tangent space of the MPS manifold of fixed
bond dimension at the current time. In this work we employ the two-site TDVP (2TDVP in Ref.~\onlinecite{Paeckel2019}) using the second order integrator by sweeping left-right-left 
with half time step $dt/2$. The main advantage of this method is that it has a smaller time-step error $O(dt^3)$, and its accuracy is controlled only by the MPS bond dimension and the threshold to terminate the Krylov series. In this work, we stop the Krylov vectors recurrence when the total contribution of two consecutive vectors to the matrix exponential is less than $10^{-12}$.\\ 
In the star-geometry considered in this work, depicted in Fig.~\ref{fig:MPS}, we have placed the qubit on the first site, on the second one the oscillator with Hilbert space of dimension $N_o$ and on the remaining sites the collection of $N$ bosonic modes of the bath, each with Hilbert space of dimension $N_{ph}$. The coupling between qubit and oscillator is $\bar{g}$, while the oscillator and the bath experiment long-range interactions with couplings $\bar{g}_j$. \\
Using the ITensor library \cite{ITensor}, we start the time-evolution from a product state with the oscillator and the bath modes in their vacuum state (zero temperature), while the qubit is placed in a generic point on the surface of the Bloch sphere. In order to reduce the simulation time and simultaneously reach a longer dynamics, we tested the $W^I$ and TDVP methods against each other on the exactly solvable closed model. We observed that the TDVP method reproduces the analytical exact solution by using a time step two or three orders of magnitude larger than that needed by $W_I$. Moreover, as expected, $W^I$ has shown a much smaller accuracy than the TDVP method. Thus, we decided to use the TDVP method for our simulations in the presence of the interaction with the bath modes. An important observation about our numerical simulations is in order: we did not use more sophisticated approaches like local basis optimization \cite{Zhang1998,Bursill1999,Friedman2000,Wong2008,Brockt2015,jansen2022finite}. We instead have converged our simulations in the number of Fock states in the oscillator and bath modes, finding the best compromise between the smallest bond dimension and longest simulation times. Our truncation error has kept below $10^{-12}$ requiring a maximum bond dimension of $D_{\rm {max}}=20$. At the same time, this optimal maximum bond dimension has allowed us to reach a time size for the simulations of $t_{\rm {final}} = 35/\Delta$, about 5 periods of the qubit. All the optimal parameters used in the MPS simulations are specified in the captions of the figures of the section \ref{sec:risultati}. A similar convergence analysis has already been done in previous works for the $W^I$ approach and for one \cite{de2020quantum} or more \cite{de2021quantum} qubits interacting with a bath. \\
We finally note that, in the star geometry, one could also adopt the TEBD method with swap gates. It was recently shown, however, in Ref.~\onlinecite{Liu2022}, that it usually requires larger bond dimensions compared to 2TDVP, despite giving smaller accumulated errors for long time evolutions.\\

In the Sect.~\ref{sec:risultati}, we compare the results of our MPS simulations against the perturbative solutions of LME and HEM. Indeed, we have used our MPS simulations to explore the parameter regions where LME or HEM are expected to fail. 
%\newpage
\appsection{Temperature effects}\label{app:temp}
In order to take into account the possible effects due to thermal fluctuations in the model, we use an Ohmic thermal bath and we analyse the weak coupling regime by means of LME. By changing the inverse temperature $\beta=1/T$ ($k_B=1$) we compute the fidelities of both the qubit (Fig.~\ref{fig:fid_temp}.a) and the oscillator (Fig.~\ref{fig:fid_temp}.b) with respect to their free evolution. These quantities reveal that the results presented in the main text at $T=0$ are robust up to temperatures of the order of $\Delta$.

\begin{figure}[h]
\centering
\includegraphics[scale=0.255]{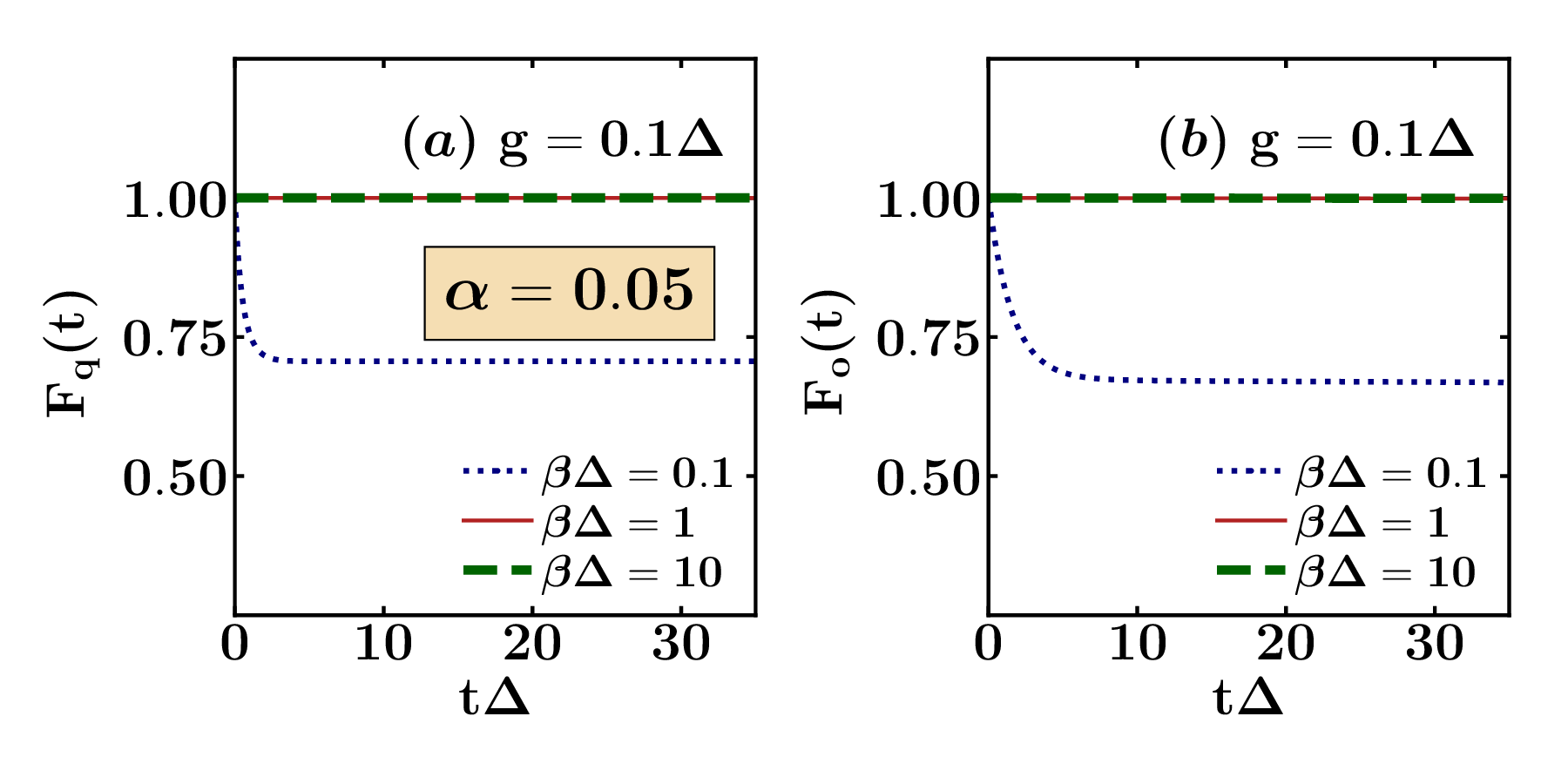}
\caption{\label{fig:fid_temp} Qubit fidelity (panel $(a)$) and oscillator fidelity (panel $(b)$) with respect to free evolution, computed at $g=0.1\Delta$ as a function of time (in units of $1/\Delta$) through Lindblad simulations for $\alpha=0.05$ and for three different values of the inverse temperature $\beta$ in units of $1/\Delta$, for the initial pure state $\ket{\nearrow}$. Parameters: $\omega_0=10\Delta,g=0.1\Delta,\alpha=0.05,\omega_c=50 \Delta,\beta=\{0.1,1,10\}/\Delta$,$N_o=20$}
\end{figure}

\end{appendices}


\bibliography{sn-bibliography}% common bib file

\end{twocolumn}

\end{document}